%% file: ms.tex
\documentclass[12pt,preprint]{aastex}

\input macros.tex

\received{}
\revised{}
\accepted{}

\shorttitle{}

\shortauthors{Jensen et al.}

\begin{document}

\title{OBSERVATIONAL PROPERTIES OF ROTATIONALLY EXCITED MOLECULAR HYDROGEN IN TRANSLUCENT LINES OF SIGHT}

\author{Adam G.~Jensen\altaffilmark{1,2,3}, Theodore P.~Snow\altaffilmark{4}, George Sonneborn\altaffilmark{1}, \& Brian L.~Rachford\altaffilmark{5}}

\altaffiltext{1}{NASA's Goddard Space Flight Center, Code 665, Greenbelt, MD 20771; Adam.Jensen@gmail.com, George.Sonneborn@nasa.gov}
\altaffiltext{2}{NASA Postdoctoral Program Fellow; fellowship administered by Oak Ridge Associated Universities}
\altaffiltext{3}{Current affiliation:  University of Maryland at College Park, CRESST}
\altaffiltext{4}{Center for Astrophysics and Space Astronomy, University of Colorado at Boulder, Campus Box 389, Boulder, CO 80309-0389; tsnow@casa.colorado.edu}
\altaffiltext{5}{Department of Physics, Embry-Riddle Aeronautical University, 3700 Willow Creek Road, Prescott, AZ 86301-3720; rachf7ac@erau.edu}

\begin{abstract}
The {\it Far Ultraviolet Spectroscopic Explorer} ({\it FUSE}) has allowed precise determinations of the column densities of molecular hydrogen ($\Hmol$) in Galactic lines of sight with a wide range of pathlengths and extinction properties.  However, survey studies of lines of sight with greater extinction have been mostly restricted to the low-$J$ states (lower total angular momentum) in which most molecular hydrogen is observed.  This paper presents a survey of column densities for the molecular hydrogen in states of greater rotational excitation ($J \geq 2$) in Galactic lines of sight with $\log{\NHmol} \gtrsim 20$.  This study is comprehensive through the highest excited state detectable in each line of sight.  $J=5$ is observed in every line of sight, and we detect $J=7$ in four lines of sight, $J=8$ in one line of sight, and vibrationally excited $\Hmol$ in two lines of sight.  We compared the apparent $b$-values and velocity offsets of the higher-$J$ states relative to the dominant low-$J$ states and we found no evidence of any trends that might provide insight into the formation of higher-$J$ $\Hmol$,  although these results are the most affected by the limits of the {\it FUSE} resolution.  We also derive excitation temperatures based on the column densities of the different states.  We confirm that at least two distinct temperatures are necessary to adequately describe these lines of sight, and that more temperatures are probably necessary.  Total $\Hmol$ column density is known to be correlated with other molecules; we explore if correlations vary as a function of $J$ for several molecules, most importantly CH and CH$^+$.  Finally, we briefly discuss interpretations of selected lines of sight by comparing them to models computed using the Meudon PDR code.
\end{abstract}

\section{INTRODUCTION AND BACKGROUND}
\label{s:intro}
$\Hmol$ is the most common molecule in the Universe, as well as a significant diagnostic of dense and dusty conditions in interstellar clouds, critical to the physics and chemistry of star-forming regions.  The many rotational-vibrational-electronic transitions of $\Hmol$ are found in the far-ultraviolet (FUV; $\lambda \lesssim 1200$ \AA{}) portion of the spectrum.  Thus, it was not until the introduction of space-based instruments with sufficient sensitivity and spectral resolving power in the FUV that detailed observations of $\Hmol$ were undertaken.  While instruments such as {\it Copernicus} were capable of measuring $\Hmol$ \citep[e.g.][]{Spitzer1973, Spitzer1974, SpitzerCochran}, observations were generally restricted to lines of sight with $\av \lesssim 1\magnitude$.  The most recent mantle-bearer of FUV---and therefore $\Hmol$---observations was the {\it Far Ultraviolet Spectroscopic Explorer} ({\it FUSE}), which operated from 1999 through 2007.  {\it FUSE} was able to push well past the extinction limits of previous FUV instruments, observing many Galactic lines of sight with $\av \sim 2-3$ and even a few targets with $\av \sim 5$.  These observations have established a somewhat surprising ubiquity of $\Hmol$, as it is found to be not just in dense, star-forming regions but is also often present in significant amounts throughout the diffuse ISM, both in the Galactic disk and at higher Galactic latitudes \citep[and references therein]{Wakker2006, Gillmon2006}.

Rotational, vibrational, and electronic splitting occurs in the energy levels of $\Hmol$.  The transitions observed in the FUV are dominated by $\Hmol$ that is in the ground electronic state, the ground vibrational state, and states of low rotational energy.  However, a detectable amount of rotationally excited, or ``high-$J$"\footnote{More technically, $J$ is the quantum number of the total angular momentum---rotational and electronic.} $\Hmol$ typically exists in the lines of sight that have been observed with {\it FUSE}.

The vast majority of the {\it Copernicus} and {\it FUSE} studies mentioned above measured either $\Hmol$ in all detected $J$ levels for lower column density lines of sight in the diffuse ISM ($\NHmol < 10^{20}$) or only $J=0-1$ $\Hmol$ in denser lines of sight ($\NHmol > 10^{20}$).  The recent survey of \citet{Sheffer2008} explored $\Hmol$ in 58 lines of sight using {\it FUSE} data, but the emphasis of that survey was on the chemical relationships between $\Hmol$ and other molecules, including $^{12}$CO, CH, CH$^{+}$, and CN.  In their paper, \citet{Sheffer2008} only indirectly reported $\Hmol$ column densities as a function of $J$ level by reporting excitation temperatures up to $T_{04}$, from which the individual column densities can be calculated (but only up to $J=4$).  Furthermore, \citet{Sheffer2008} did not address issues specific to the excitation of the various $J$ levels.

In contrast, the aim of this paper is to carry out a survey inclusive of all detectable $J$ levels and any detectable vibrationally excited material.  (In our survey this turns out to be at least $J=5$ in every line of sight; see \S\ref{s:results} and Table \ref{coldensities}.)  We will do this by examining reddened lines of sight from the {\it FUSE} archives, primarily from a series of papers \citep{Rachford2001, Rachford2002, Rachford2009, Snow2000HD73882, Sonnentrucker2002}, many of which were led by authors on this paper (BLR and TPS), that used {\it FUSE} to examine lines of sight thought to potentially contain translucent clouds.  We will hereafter refer to this collection of papers, with emphasis on the two papers \citep{Rachford2002, Rachford2009} that were surveys of $J=0-1$ $\Hmol$ in these lines of sight, as the {\it FUSE} Translucent Cloud Survey, or FTCS.  Our intent is to supplement the measurements of $N(0)$ and $N(1)$\footnote{Our notation from this point forward is that $N(J)$ refers to the column density of $\Hmol$ in the $J$ state (or states if a range of $J$ is specified).  The alternate, condensed notation $N_J$ is used in our figures and tables.} in the FTCS survey papers with measurements of $N(\geq 2)$.  We will loosely call $\Hmol$ in the $J \geq 2$ states ``rotationally excited" or ``high-$J$" at points in this paper, although a better physical distinction occurs at about $J=3$ (see \S\ref{ss:Texc}).  The target sample is discussed more in \S\ref{s:obsdata}.

There are several issues that our study intends to address.  First, any study based on a data set of reddened lines of sight may help better characterize the nature of translucent clouds.  Qualitatively, translucent clouds are a transition phase between diffuse and dense clouds in the ISM.  In the diffuse ISM, the interstellar material is dominated by neutral atoms and ions, hydrogen is nearly all atomic, dust is present but gas dominates the material, and ices do not form on grains.  In contrast, in the dense ISM, clouds are dominated by molecules (including molecular hydrogen), the depletion of atoms directly onto dust grains is significant, and ices readily form on grains.  In translucent clouds, a transition between these properties is expected to be observed---atoms and molecules will have comparable abundances and chemical significance, the molecular fraction of hydrogen [$\fHmol \equiv 2\NHmol/(\NHI + 2\NHmol)$] will approach unity, and ice formation on dust grains begins.  In addition to the FTCS papers and references therein, recent discussion of translucent clouds can be found in \citet{SnowMcCall}; in particular, note Fig.~1 of that paper.

In this context of translucent lines of sight, we can further address simple first-order questions about higher-$J$ $\Hmol$ in the ISM.  For example, preliminary work on this topic \citep{RachfordBakerSnow2002} suggested that higher-$J$ $\Hmol$ might exist in a line of sight primarily in a warmer shell around the colder cloud core containing the lower-$J$ $\Hmol$.  Another simplistic possibility is that there two distinct clouds along a line of sight, one cold and one warm.  The hypothesis that higher- and lower-$J$ $\Hmol$ are not well-mixed can be tested in part by analyzing the $b$-values (the velocity dispersion if a single distribution is assumed) and velocity offsets of the different $J$ levels---differences in either or both would provide circumstantial evidence that the low-$J$ and high-$J$ $\Hmol$ is, at least in part, physically distinct.  For example, if there is detectable, systematic offset in velocity offsets between high- and low-$J$ $\Hmol$ in a given line of sight, that might indicate a situation similar to or approximated by the two-cloud case.  A distribution of such offsets or differing $b$-values over many lines of sight that cannot be explained as statistical variation could indicate the degree to which the different $J$-levels are mixed.  Note that a null result, however, does not necessarily indicate the converse, that high- and low-$J$ $\Hmol$ are well-mixed in most or all cases.

Furthermore, the question of how higher-$J$ $\Hmol$ forms and is sustained in the ISM remains open.  \citet{Nehme2008} examined the {\it FUSE} spectrum of HD 102065, and attempted to model the line of sight with the Meudon PDR code \citep{LePetit2006}.  \citeauthor{Nehme2008} found that the Meudon PDR code could reproduce some of the basic observational properties of this line of sight, such as the low-$J$ column densities (including the ratio of ortho- to parahydrogen) and the column densities and emissivities of certain molecules.  However, there are other properties that are not explained, such as the column densities of $J \geq 3$ $\Hmol$ and CH$^+$. 

The scope of this paper is to present measurements of $N(\geq 2)$ in 22 selected lines of sight from the FTCS.  We will also present some interpretations and simple modeling of our results, using the Meudon PDR code.

\section{OBSERVATIONS AND DATA REDUCTION}
\label{s:obsdata}
The spectra under examination are taken from the {\it FUSE} archives.  As noted above, our primary source for lines of sight under consideration was the FTCS.  This comprises a database of about 40 lines of sight.  From this database, we have selected a sample of 22 lines of sight.  Our first selection criterion was a signal-to-noise ratio (S/N) large enough to detect the weaker lines of $\Hmol$ in the $J=4-5$ states, combined with a qualitative assessment of the continuum that would allow for the simple measurement of most lines.  Our second selection criterion was that the line of sight's $\Hmol$ content had not been previously examined in detail.  This latter criterion eliminated five lines of sight (of the original 40) from consideration---HD 73882, HD 96675, HD 102065, HD 108927, and HD 110432.  HD 73882 and HD 110432 were analyzed as part of the FTCS \citep[respectively]{Snow2000HD73882, Rachford2001}, while the other three lines of sight were analyzed in \citet{Gry2002}.  Furthermore, four of these lines of sight (all but HD 73882) have had their column densities used as the subjects of modeling---HD 102065 in \citet{Nehme2008} and the other three in \citet{BTS2003}.

Of the lines of sight that we do include, a few have been analyzed elsewhere, notably HD 185418 in \citet{Sonnentrucker2003}, HD 192639 in \citet{Sonnentrucker2002} and \citet{Lacour2005}, and HD 206267 and HD 207538 also in \citet{Lacour2005}.  However, we have reanalyzed these lines of sight for various reasons.  Our fitting method was different than \citet{Sonnentrucker2002, Sonnentrucker2003} and we detect $J=6$ in HD 192639 whereas \citet{Sonnentrucker2002} do not report a detection.  Furthermore, the analysis of the higher $J$ levels is not discussed to the same degree in these papers as it is in the papers examining the excluded lines of sight mentioned above.  The reason for reanalyzing the \citet{Lacour2005} lines of sight is that our methodology is different as we are using additional curve-of-growth methods to explore various possibilities for the velocity structure.  The lines of sight that are included in our study are shown in Table \ref{stellardata} along with some basic line of sight parameters.

\subsection{{\it FUSE} Data}
\label{ss:FUSEdata}
The {\it FUSE} satellite recorded data on four different physical detectors (LiF1, LiF2, SiC1, and SiC2), named after the two different reflecting materials on the optical elements---lithium fluoride and silicon carbide.  Each detector has two adjacent segments, covering different wavelengths, denoted A and B (e.g.~LiF1A and LiF1B).  For further information on {\it FUSE} instrumental details and performance, see \citet{Moos2000} or \citet{Sahnow}.

The spectra were processed in the CALFUSE pipeline, using version 2.4.0 or later.  All of the analysis described below was conducted in different detector segments independently, before merging results from various segments.  This was done to avoid the possible effects of any instrumental features because, unlike a study using a limited number of absorption lines (e.g.~\citeauthor{JensenFeII}~\citeyear{JensenFeII}), the lines in this study cover nearly the entire {\it FUSE} wavelength region.  Therefore, a global coadd of different detector segments is sure to include at least some detector segment-specific instrumental features.  Analyzing the detector segments individually, therefore, provides us with the ability to look note possible instrumental effects.  Some minor unidentified features were noted that influenced whether or not some absorption lines were included, but this affected, at most, one or two lines per $J$ level.

\subsection{Auxiliary Data}
\label{ss:auxdata}
In order to interpret our results, it is useful to have several pieces of auxiliary information, including but not limited to $N(0)$ and $N(1)$.  These low-$J$ column densities of $\Hmol$ are taken from the FTCS, except for HD 195965, the only line of sight in our sample which is not in the FTCS.  For this line of sight, $N(0)$ and $N(1)$ are taken from a {\it FUSE} survey of $\Hmol$ in the Galactic disk (J.~M.~Shull 2010, in preparation).  Additional observational values for these lines of sight, such as molecular column densities and values of reddening and extinction, are taken from various sources.  Reddening and extinction values, including references, are given in Table \ref{reddeningtable}.

There are literally hundreds of $\Hmol$ transitions which have wavelengths corresponding to the spectral coverage of {\it FUSE}.  In this study, we typical measure $\sim80$ lines of $J \geq 2$ $\Hmol$ per line of sight (though more may be observed; see \S\ref{ss:H2lines}).  Wavelengths and oscillator strengths for the $\Hmol$ absorption lines were taken from \citet{Abgrall1993a, Abgrall1993b}.  These data were used to identify observed transitions, derive velocity offsets, and make curve-of-growth fits (see \S\ref{ss:H2columns}).

\subsection{Fitting of $\Hmol$ Lines}
\label{ss:H2lines}
We follow our previously established methods, most recently outlined in \citet{JensenFeII}.  Our goal is to measure equivalent widths for use in curve-of-growth determinations of column densities.  This is detailed below in \S\ref{ss:H2columns}.  To summarize, we measure the equivalent widths of observed absorption lines by making simple fits to the local continuum and structure of the lines.  The local continuum around each line is fit with a low-order polynomial.  Weaker absorption lines, where the observed profile is dominated by the instrumental profile, are fit with Gaussians.  Absorption lines where the observed profile depth approaches unity are fit with generalized Voigt profiles, assuming a $15\kmpers$ full-width at half-maximum (FWHM) Gaussian instrumental profile for {\it FUSE}.  Errors are derived from standard error propagation in the Gaussian case, including continuum placement.  In the Voigt case, errors are dominated by continuum placement errors, and an estimate of this error is obtained by assuming a 1-$\sigma$ error in continuum placement, and then integrating this value over the wavelength range where the line profile is at a nontrivial depth ($> 1\%$, a more than sufficient criterion for the deep lines under consideration).  These two methods of error estimation are observed to be comparable when both are used on test cases.

In most lines of sight, there is little to no evidence that the absorption profiles deviate significantly from a single component absorption structure, though the moderate resolution of {\it FUSE} (approximately $R=20,000$) masks the true component structure.\footnote{Note that a typical thermal width of $\Hmol$ should be $\sim0.8\kmpers$ and $\sim1.5\kmpers$ for lower- and higher-$J$ $\Hmol$, respectively, based on the temperatures we derive in \S\ref{ss:Texc}.  This calculation assumes a single cloud with no turbulence.  These $b$-values are significantly smaller than the {\it FUSE} resolution, which translates to a $b$-value of $\sim9\kmpers$.  However, turbulent broadening or a multiple cloud structure may result in an observed $b$-value that is comparable to the resolution.}  However, in five lines of sight (HD 40893, HD 149404, HD 195965, HD 1999579, and HD 210839), there are consistently asymmetric profiles, which demand that we use a fit with multiple components.  In these cases, we extend the methods described above to multiple components and record the total equivalent width.  With this method we obtain an independent equivalent width for each component, but the equivalent width we use in the curve of growth is the total resulting equivalent width of the entire profile, not a linear sum of the two equivalent widths.  We never find a consistent need for more than two components in these lines of sight that require multiple components.  In the five lines of sight where multiple components are identified, we use the same curve-of-growth methods described below (\S\ref{ss:H2columns}), which employ varying assumptions.  Table \ref{eqwidths} lists all the equivalent widths and 1-$\sigma$ errors.

The absorption lines of $\Hmol$ are occasionally blended with other lines, both other $\Hmol$ lines and atomic lines.  The comparison list of atomic lines is from \citet{Morton2003}; HD lines are also considered.  In these cases, our minimum criterion for proceeding with a line measurement is if the line centers are clearly resolved.  In these cases, we simultaneously fit the partially blended lines and record the equivalent width fits.  In cases where the line centers are unresolved or multiple components significantly complicate the fitting, we exclude these transitions from our analysis.

In addition to obvious blending from stronger atomic lines or other observed $\Hmol$ lines, there are many cases where the overlap between two $\Hmol$ lines is nearly complete, such that it is not visually obvious that the feature is the blend of multiple lines.  However, if the feature is clearly dominated by one line (e.g.~a strong $J=2$ line with a $J=7$ line in a line of sight where $J=7$ is otherwise undetected), it still may be appropriate for inclusion in our study.  We determine the suitability of such lines by examining unblended lines (if they are detected) of the same $J$ level of the line which is presumed to be weaker.  Generally, if the presumed weaker line is from a state of $J \leq 6$, the stronger line is unsuitable for inclusion.  If $J \geq 7$, then the suitability of the stronger line is dependent on that particular line of sight.

\begin{figure}[t!]
\begin{center}
\epsscale{1.00}
\plotone{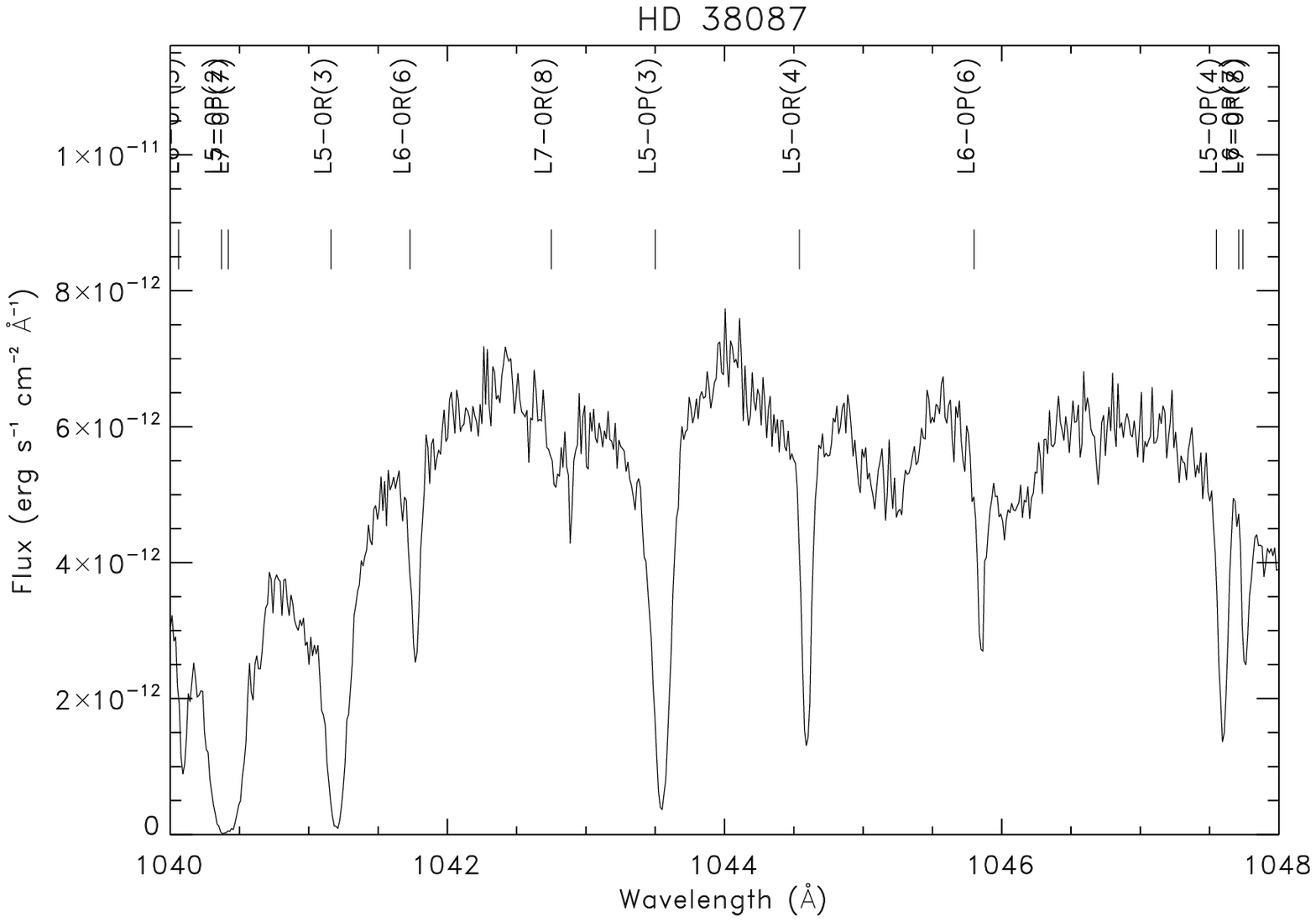}
\end{center}
\caption[Sample spectra of HD 38087.]{Above is an interesting region of the {\it FUSE} spectrum of HD 38087, showing $\Hmol$ lines up to $J=8$.  The tick marks are at laboratory wavelengths; in this spectrum, the lines are shifted by $\sim11\kmpers$ ($< 0.5$ \AA{}).  $J$ levels are in parentheses.  HD 38087 shows the only clear detections of $J=8$ in our sample of 22 lines of sight; the detected $J=8$ line here is bluer feature of the two features near 1042.8 \AA{}.}
\label{fig:specsample}
\end{figure}

A sample of absorption lines in HD 38087 is shown in Fig.~\ref{fig:specsample}.  This line of sight shows the greatest $\Hmol$ excitation in our sample, as we detect up to $J=8$.

\subsection{Column Density Limits and Vibrationally Excited $\Hmol$}
We also briefly explored the limits of the next undetected rotationally excited $J$ level and searched for vibrationally excited $\Hmol$.  Our upper limits are based on the equation
\begin{equation}\label{eq:eqwlimit}
W_{\lambda,max}=\frac{N_{\sigma} d\lambda \sqrt{M}}{\rm S/N}
\end{equation}
where $W_{\lambda,max}$ is the upper limit on $\eqw$, $N_{\sigma}$ is the number of $\sigma$ confidence desired, $d\lambda$ is the wavelength spacing of the pixels, $M$ is the number of pixels required to scan $\eqw$, and S/N is the signal-to-noise of the local continuum.  We assumed 15 pixels for detection (the {\it FUSE} FWHM is 9 pixels corresponding to a $b$-value of $\sim9\kmpers$), and calculated 1-$\sigma$ upper limits on $\eqw$; column density limits can then be calculated from matching the limit on $\eqw$ to our preferred curve of growth solutions (see \S\ref{sss:Nconclusions}).

While investigating these limits, we also detected a few lines of vibrationally excited $\Hmol$ in HD 38087 and HD 199579.  The strengths and velocity offsets of these lines are consistent with our observations of the other $\Hmol$ excitation levels.  Detections of vibrationally excited $\Hmol$ in cold diffuse clouds are very rare, although at least two lines of sight from the FTCS sample that are not included in this study show $\Hmol$ in several of these levels---HD 37903 \citep[discussed, along with a review of the topic, in][]{Meyer2001} and HD 164740 (B.~L.~Rachford et al.~2010, in prep.).  However, because we only detected a total of five absorption lines in two lines of sight in this sample, we do not include these detections in the rest of our main analysis (\S\ref{s:results}).

\subsection{Derivation of High-$J$ $\Hmol$ Column Densities}
\label{ss:H2columns}
Few, if any, of the high-$J$ absorption lines that we examined in this study are of an appropriate strength\footnote{By ``strength" we essentially mean the equivalent width, $\eqw$, which results from a combination of the line of sight properties (column density and velocity structure) and the inherent properties of the transition (oscillator strength $f$ and damping constant $\gamma$).} that a reliable column density can be determined directly from a single line.  Rather, nearly all of the lines have at least some inherent saturation, but are not so strong as to show the clearly damped profiles of the $J=0-1$ lines (or allow for column densities to be accurately determined through the $\eqw \propto \sqrt{N}$ relationship that exists in this regime).  Therefore, we use a curve-of-growth method to derive column densities.  We use three versions of the curve-of-growth method in an attempt to be complete.

The first method is to fit the lines of each $J$ level of $\Hmol$ to separate single-velocity dispersion curves of growth.  Each $J$ level is allowed to vary in column density and $b$-value.  The column density range is $\log{N}=10-22$ in increments of $0.01\dex$ and the $b$-value range is $b=0.1-30\kmpers$ in increments of $0.1\kmpers$, resulting in a $1201\times300$ grid of possible solutions in ($N$, $b$).\footnote{It is worth noting that the ranges on $N$ and $b$ are much broader than the expected solutions, by $\gtrsim2\dex$ for each limit on $N$, and a factor of $\gtrsim2$ for the upper limit on $b$.}  This method avoids some assumptions that may not be justified, e.g.~that different $J$-levels have the same $b$-value, or that the velocity structure precisely matches the structure of some other atom or molecule.  However, the uncertainties inherent to this method can be very large, both the statistical uncertainties when most of the absorption lines are on the ``flat" portion of the curve of growth, and the systematic uncertainties if the velocity structure is not closely approximated by a single-velocity dispersion.  As a result, when we determine the error ellipse by calculating $\chi^2$ at each point in the column density/$b$-value grid, the resulting error is sometimes very large as well.  In addition, we sometimes see unusual solutions (e.g.~population inversions that are extremely unlikely to be real) that imply systematic errors.

The second method is similar to the method described above, but instead we fit all the $J$ levels of $\Hmol$ to the same single-velocity dispersion curve of growth, instead of allowing for independent $b$-values for each $J$ level.  This method helps to constrain the column density solutions, but whether or not this assumption accurately describes the physical conditions is uncertain.  At the least, it prevents us from exploring one of the questions mentioned in \S\ref{s:intro}, concerning whether or not there is a difference in the $b$-value of low- vs.~high-$J$ $\Hmol$.

The third method uses multiple-velocity dispersion curves of growth based on the velocity structure of appropriate molecules.  Specifically, we use the column densities and velocity structure of CH and CH$^+$ that have been measured by D.~E.~Welty (2008, private communication).  This method is physically justified in that $\Hmol$ and CH are known to be strongly correlated (see the FTCS); furthermore, higher-$J$ $\Hmol$ and CH$^+$ are both known to be correlated \citep{LambertDanks}.  We note that radiative transfer models \citep[e.g.][]{Nehme2008} are unable to explain observations of the high-$J$ $\Hmol$ and CH$^{+}$.  Like the previous method of assuming a uniform $b$-value, this method prevents us from directly examining whether or not low- and high-$J$ deviate from full physical coincidence.  Rather, this method makes an implicit assumption about whether this is the case, based on which $J$ levels are fit to which species and whether CH and CH$^+$ are physically coincident or distinct.

For all of these methods, the choice of a damping constant ($\gamma$) affects the fit---this is virtually always the case for $J=2$, occasionally the case for $J=3$, and virtually never the case for $J \geq 4$.  Therefore, in our $\chi^2$ calculation, the difference between the measured data points and the theoretical curves are based on different curves, appropriate to the damping constant of the absorption line under consideration.  For example, the $J=2$ line at 1081.267 \AA{} has a damping constant of $\gamma_{1081.267}=1.63\times10^9 {\rm \;s}^{-1}$, while the 1016.458 \AA{} line has a damping constant of $\gamma_{1016.458}=1.236\times10^9 {\rm \;s}^{-1}$.  Therefore, when calculating the value of $\chi^2$ for a column density of $\log{N}=18.50$ and $b=2.0\kmpers$, the difference terms in the $\chi^2$ sum use curves of growth with different values of $\gamma$, but the same $b$-value.

In order to simplify the computational process slightly, not every possible damping constant was calculated.  The damping constants of the lines under considerations are bounded by $5\times10^8 {\rm \;s}^{-1} < \gamma < 2\times10^9 {\rm \;s}^{-1}$; therefore, theoretical curves of growth were calculated for $\gamma$ on this range, in increments of $5\times10^7 {\rm \;s}^{-1}$.  In the worst case, these values of $\gamma$ are separated by 10\%, meaning that each absorption line will be matched to a value of $\gamma$ within 5\% of its true value of $\gamma$.  In the worst case scenario where lines are fully damped, $\eqw \propto \sqrt{\gamma}$; therefore, this method is equivalent to introducing as much as a 2.5\% error in $\eqw$.\footnote{It should be noted, conversely, that $\eqw \propto \sqrt{N}$ on this regime, so the error in $N$ would be as high as 5\% if based on a single line.}  However, we note the following three mitigating factors:  (1) the errors should be more or less randomly distributed over the dozens of lines measured, resulting in a reduced cumulative effect, (2) the introduced error is usually much smaller than this (it is negligible for the weaker lines), and (3) even in this worst case, this error is comparable to the statistical fitting errors on the measured values of $\eqw$, and is usually smaller.

\section{RESULTS}
\label{s:results}
In this section, we will discuss the results of the first level of analysis based on the line measurements.  First, we compare the various methods of deriving column densities.  Then we will turn our attention to understanding the $b$-values and velocity offsets of the absorption lines, and what this says about the nature of where higher-$J$ $\Hmol$ is formed and/or exists.  In the following section (\S\ref{s:discussion}) we will use our adopted column densities to explore the additional issues of correlations with other atoms and molecular, excitation temperature, and radiative transfer modeling these lines of sight.

\subsection{Column Densities}
\label{ss:coldensities}
As discussed above, we performed column density analyses in three different ways---fitting column density and $b$-value independently for each $J$ level, fitting all $J$ levels to a single $b$-value (and independent column densities), and fitting all $J$ levels to a predetermined curve of growth.  Making quantitative comparisons between these methods is difficult because of the systematic uncertainties.  Nevertheless, we must attempt to understand which of these methods produces the most accurate results.

\subsubsection{Independent $b$-values}
\label{sss:indbvalues}
When we assume that $b$-values are independent, we notice, in general, trends of increasing $b$ with increasing $J$.  Compared to a uniform $b$-value, this will tend to decrease the column densities of the higher-$J$ $\Hmol$.  It is important to note that in some but not all cases this effect is negligible because the error in $b$ is large---the best fit is simply finding that the higher-$J$ $\Hmol$ appears to be on the linear portion of the curve of growth, and the $b$-value fit and its error should instead be interpreted as a lower limit.

We have attempted to look for direct causes of potential systematic uncertainties by examining anomalous absorption lines.  There are certain lines which frequently have equivalent widths that are much larger or smaller than expected; lines which are systematically stronger or weaker than expected across multiple lines of sight are consistently identified as having a continuum that is less certain than other lines.  When we exclude these lines that have apparent systematic uncertainties from the curve-of-growth analysis, the curve-of-growth fits are improved in many cases.  However, even after this consideration there are still some column density results that we consider anomalous---specifically, strong deviations from a reasonable population expectation, that is, population inversions isolated to one $J$ level.  Possible additional systematic uncertainties that we have not considered include undetected blending with other atomic or molecular transitions or $f$-value errors.

Our conclusion is that this method, while preferable aesthetically because of its lack of {\it a priori} assumptions, is ultimately the least reliable, because there are fewer constraints, namely a much smaller range in $f\lambda$.  This enhances the potential for systematic errors in the final column density derivation due to inherent deviations from a single-component curve of growth or unrealized systematic errors in the measurements of the absorption lines.

\subsubsection{Uniform $b$-value}
\label{sss:uniformb}
Assuming a uniform $b$-value for all $J$ levels of $\Hmol$ is a good way to put constraints on the curve of growth method and obtain a seemingly reasonable solution.  However, while a uniform $b$-value seems like a good first-order assumption, there is a distinct possibility that this may not be the case.  Discussion of this phenomenon can be found in earlier {\it Copernicus} results, such as \citet{Spitzer1974}.  Papers by \citet{Jenkins1989} and \citet{JenkinsPeimbert} also found some evidence that $b$-value varies as a function of $J$---although the nature of the effect was the opposite for the two lines of sight in those papers (decreasing $b$-value with increasing $J$ the former paper's examination of $\pi$ Sco, increasing $b$-value with increasing $J$ in the latter paper's discussion of $\zeta$ Ori A).  A more recent example can be found in \citet{Lacour2005}, who determined to 4-$\sigma$ confidence that HD 192639 cannot be fit by a single $b$-value---$b$ increasing with increasing $J$ in this case.  In their paper, \citeauthor{Lacour2006} also discuss the implications this has for the formation mechanisms of $\Hmol$.  This line of sight is also in our sample, so we are able to perform the analysis using both methods (independent and uniform $b$-value) on this line of sight.  When fitting $b$-values independently to each $J$ level, we do not rule out a single $b$-value at the same high confidence as \citet{Lacour2005} but our results are qualitatively similar.  In addition, quantitative differences between our column densities and those of \citet{Lacour2005} are observed at levels between 1- and 3-$\sigma$ for $J=2-4$.

However, it may still be useful to attempt to fit all $J$ levels to a single $b$-value.  As a practical matter, this helps to restrict $b$-value solutions to qualitatively reasonable values (i.e.~no population inversions), but may introduce systematic errors if this assumption is invalid.  Furthermore, in spite of the aforementioned potential evidence to the contrary in \citet{Lacour2005}, assuming a single $b$-value is the simplest {\it a priori} assumption, and is worth more exploration.

Even within the context of assuming a ``uniform" $b$-value, we might still find multiple $b$-value regimes as a function of $J$.  What if, for example, $\Hmol$ in the $J \le 3$ states is at one $b$-value, while $\Hmol$ in the $J \ge 4$ states is at another?  This would be consistent with the observation of different excitation temperature regimes as a function of $J$ (see \S\ref{ss:Texc}).  Unfortunately, we cannot probe the $b$-values of $J=0-1$ because these lines are strongly damped and therefore the profiles are less sensitive to changes in $b$-value.  Therefore, between continuum uncertainties and the $\sim15\kmpers$ resolution of {\it FUSE}, profile fitting cannot easily differentiate between $b$-values for these lines.  Curve-of-growth methods likewise cannot easily distinguish between $b$-values, with all the lines being well onto the damped portion of a typical curve.  This is also potentially a problem for the $J=2$ lines, which largely exist at the flat-to-damped transition and are only mildly sensitive to $b$-value---however, with the stronger $J=2$ lines setting the column density, the weaker $J=2$ lines are frequently able to help set the $b$-value with reasonably high confidence.

To explore this, we performed three variations of the ``uniform $b$-value" method:
\begin{enumerate}
\item Assume that $b$ is uniform for $J \geq 2$ ($b_{2+}$)
\item Assume that $b$ is uniform for $J \geq 3$ but independent for $J=2$ ($b_{3+}$ and $b_2$, respectively)
\item Assume that $b$ is uniform for $J \geq 4$ and an independent $b$-value is shared by $J=2-3$ ($b_{4+}$ and $b_{23}$, respectively)
\end{enumerate}

First, we note that there is a great deal of consistency for $J \geq 4$ for all three methods---16 of the 22 lines of sight match within their 1-$\sigma$ column density errors.  An additional three lines of sight (HD 27778, HD 179406, and HD 206267) match within 1-$\sigma$ errors for $J \geq 5$ column densities, but have as much as a $0.04\dex$ gap between the largest lower limit and the smallest upper limit for $J=4$.  The remaining lines of sight are HD 24534, HD 53367, and HD 210839.  The discrepancies for HD 210839 are larger than the errors but not extreme:  $0.16\dex$ for $J=4$ and $0.06\dex$ for $J=5$.  For HD 24534 and HD 53367, all of the discrepancies are at least $0.24\dex$, and are greater than $1\dex$ for the two HD 24534 column densities.  In these cases, methods 1 and 2 are consistent, but method 3 sets values of $b_{4+}$ and $b_{23}$ that are significantly different both from each other and from the $b$-values derived by methods 1 and 2.

In general, the biggest variations are for $J=3$.  While several of our lines of sight are either consistent within their errors or have minimal variation (5 agree within errors; another 3 agree within errors plus 0.04 $\dex$), the majority of lines of sight are not this close, with disagreements of over 1 $\dex$ in many cases.  In 20 of 22 lines of sight, method 3 is the outlier of the three methods, while methods 1 and 2 are consistent within errors for 15 lines of sight, and consistent within errors plus 0.1 $\dex$ for another five lines of sight.

The nature of these discrepancies makes sense because the $J=2$ equivalent widths, nearing the damped portion of the curve of growth, are less sensitive to $b$-value and therefore do not constrain it as much in the method 3 fit; the $J=3$ lines are then free to determine the $b$-value in the fit, and suffer from the systematic uncertainties similar to the independent $b$-value case.  However, in methods 1 and 2, where the $J=3$ equivalent widths are ``linked" to the $J \geq 4$ equivalent widths, the $J=3$ column density fits are constrained to a better-justified $b$-value.  These qualities cause us to rule out method 3 compared to the other two methods.

There is also some variation in the $J=2$ fits between the three methods, greater than the variations in $J=4-5$ but not as much as for $J=3$; again, this makes sense because the $J=2$ column density fits should be less sensitive to $b$-value.  Comparing methods 1 and 2, where $J=2$ is linked with the other column densities versus allowed to be fit to its own $b$-value, we see consistency within the errors in the majority of cases (12 of 22).  Another six cases have disagreements of $0.07\dex$ or less (typically 2- or 3-$\sigma$ discrepancies).  In the remaining cases, a qualitative check favors method 1---adopting method 2 would imply a very small ratio of $J=2$ to $J=3$, to the point of having a negative excitation temperature, $T_{23}$ in two of the four cases (see also \S\ref{ss:Texc} for more general discussion of excitation temperatures).

The consistency check between these variations gives a sense of what potential systematic errors are inherent to the different methods.  We have concluded that method 1 is the best version of these three variations, based on the simplicity of its underlying assumption and the avoidance of unusual solutions (e.g.~a negative $T_{23}$).  We note, however, that in principle any individual line of sight might be better fit by a different variation.  Further discussion concerning the best choice of curve-of-growth methods is found in \S\ref{sss:Nconclusions}.

\subsubsection{Predetermined Curves of Growth}
\label{sss:predetermined}
We calculate curves of growth from fits of CH and CH$^+$ column densities made by D.~E.~Welty (2008, private communication) using various high-resolution ($1.2-3.6\kmpers$) ground-based optical data sets\footnote{Information about the majority of the data can be found at http://astro.uchicago.edu/$\sim$welty/.}, and then fit the $\Hmol$ lines to these curves.  In creating these curves of growth, the $b$-values have been scaled with the mass of the molecule, according to the formula:
\begin{equation}\label{eq:bvalue}
b^2 = \frac{2kT}{m} + \xi^2
\end{equation}
In this equation, $m$ is the mass of the molecule and $\xi$ is the turbulent velocity.  Because the turbulent velocity and temperature are not explicitly known, one or the other must be assumed.  Therefore for each identified cloud component of CH or CH$^+$ with a measured $b$-value, we assume a temperature, then use the $b$-value and the molecular mass to solve for $\xi$.  We then use that value of $\xi$ and the $\Hmol$ mass to calculate the new assumed $b$-value for $\Hmol$ in that cloud.

The temperature that is assumed is 70 K for all lines of sight, which is broadly consistent with the average core temperature that was derived in the FTCS and our study (see \S\ref{ss:Texc}) through measurement of the low-$J$ excitation.  However, the variations in $b$-value due to inaccuracies in this temperature assumption are small---$b$ varies as $\sim\sqrt{T}$ for small $\xi$, and as $\xi$ becomes larger, small differences in temperature have a reduced effect.  In our data, $\xi$ usually dominates the $b$-value, i.e., $\xi$ is usually greater than $\sqrt{\frac{2kT}{m}}$.  Therefore, universally assuming a temperature of 70 K should result in a reasonable scaling of the $b$-value.  Attempting to further model the temperature will not result in significantly increased precision due to the inherent uncertainties of the assumptions that are being made.  Another issue is that the higher $J$ levels are found at a higher excitation temperature (again see \ref{ss:Texc}), but this may or may not translate to the thermal temperature because the higher $J$ may be partially influenced by radiation.  With the various combinations of CH, CH$^+$, and lower- and higher-$J$ $\Hmol$, we would potentially have to assume multiple temperatures with little observational guidance.  We also note again that when $\xi>\sqrt{\frac{2kT}{m}}$, even a factor of a few in temperature may not have a significant impact on the scaled $b$-value.  However, scaling the $b$-value in this manner with the assumption of a 70 K temperature should provide an improvement compared to no scaling.

The foremost question is whether to use the CH or the CH$^+$ data.  As discussed above, total $\Hmol$ column density [dominated by $N(0)$ and $N(1)$] is generally correlated with CH and higher-$J$ $\Hmol$ is correlated with CH$^+$.  However, it would be prudent for us to confirm these correlations in our data set, and also note that correlations do not necessarily mean that assuming a common velocity structure is the most appropriate curve-of-growth method.

Attempting to both look for correlations and assume a correlation so that a curve of growth can be assumed involves obvious circular reasoning.  Below in \S\ref{ss:molecules} we have examined the different $J$-levels for correlations with CH, CH$^+$, and other atomic and molecular species.  We find that some of the most consistent correlations are with CH and CH$^+$.  This was done for the $\Hmol$ column densities determined assuming a uniform $b$-value.  We do confirm a high level of correlation for both CH at very low-$J$ and CH$^+$ with $J=3-5$.  However, neither is correlated at statistically significant levels for $J=2$ or $J\geq6$.  It is also worth noting that we only have velocity structure data for 17 of our 22 lines of sight, and both CH and CH$^+$ data for only 14 lines of sight (note that there are additional column density derivations, which are based on equivalent width measurements but not a full fit to the component structure, that are available and included in our analysis in \S\ref{ss:molecules}).  Therefore, if we were to adopt this method, we would not be able to perform a uniform data analysis over the whole set.

\subsubsection{Column Density Conclusions}
\label{sss:Nconclusions}
\begin{figure}[t!]
\begin{center}
\epsscale{1.00}
\plotone{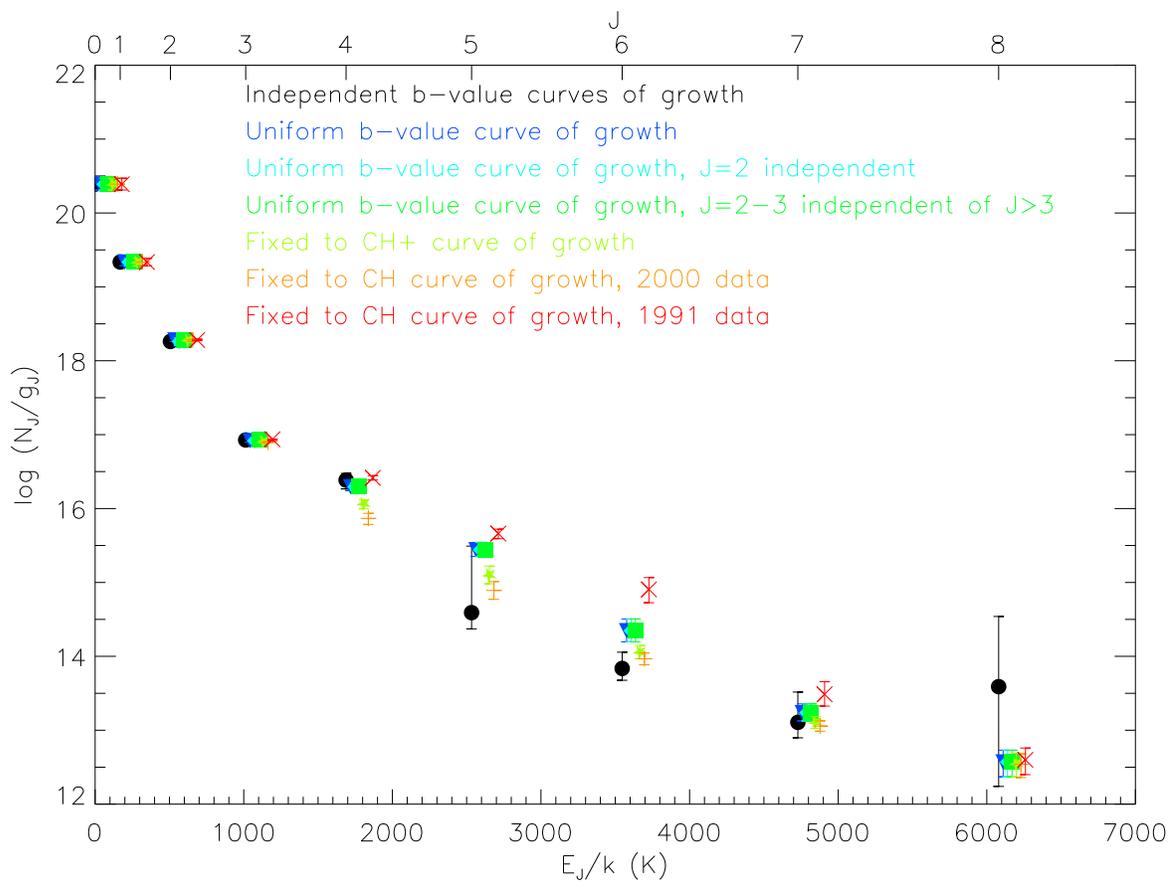}
\end{center}
\caption[HD 38087 Curve-of-Growth Variations.]{Column density results of several different variations of the curve-of-growth method for HD 38087.  Brief color-coded descriptions are given in the plot; see text for further details.  Slight offsets along the abscissa are introduced for clarity.  $N(0)$ and $N(1)$ are not independently derived, and thus are the same for all methods.}
\label{fig:COGvariations}
\end{figure}

In Fig.~\ref{fig:COGvariations} we show an example of how variable the different methods are.  This is done for HD 38087.  In this line of sight, $N(2)$ and $N(3)$ are completely consistent within their errors [note that $N(0)$ and $N(1)$ are shown in the plot but not independently derived].  From $N(4)$ through $N(6)$, and to a lesser degree $N(7)$, there is some disagreement between the methods; at $N(8)$ the methods are once again consistent (noting the very large error on the independent $b$-value derivation for $N(8)$, because the limited number of absorption lines are consistent with a wide range of $b$-values using the curve of growth method).  The inconsistencies are obviously due to the differing levels of saturation that is implied by each method.  What may be initially perplexing is that the column densities based on the CH velocity structure measured in two different data sets have the most variation (again when the large errors of the independent $b$-value results are considered).  This is explained by the inconsistency between the 2000 and 1991 data sets for CH.  The fit $b$-value of the single observed CH component in these two data sets differs by a factor of nearly two; however, the fitting errors in the 2000 data set are much smaller ($2.2\pm0.4\kmpers$ compared to $1.3\pm1.8\kmpers$), because the 1991 data set is particularly noisy (D.~E.~Welty 2009, private communication).  The much smaller $b$-value implies that the moderate-strength absorption lines of the $J=4-7$ are much more saturated, hence the larger resulting column densities.  Because our fixed curves of growth are calculated without consideration of the error in the fits in these predetermined $b$-values, we cannot account for uncertainties such as this in any way that is not computationally and conceptually complex.  It is somewhat notable how consistent the independent $b$-value method is with the variations on the uniform $b$-value method within their errors, with the exception of $N(6)$.

Overall, we find a fair amount of consistency between the three methods.  The more consistent methods are assuming a uniform $b$-value and using predetermined curves of growth---a signal that using a uniform $b$-value is frequently a fair approximation of the true velocity structure.  However, there are occasionally significant discrepancies both between the three methods (independent $b$-value, uniform $b$-value, and fixed velocity structure), and within their variations (assuming a ``break" in the $b$-value of the uniform $b$ method or different fixed curve-of-growth component structures).

We have elected to use the results from the uniform $b$-value method, assuming that the $b$-value is constant across all $J$ levels.  This is a decision that is equal parts quantitative and qualitative.  Despite the possibility that $b$-values may vary as a function of $J$ (see earlier discussion in \S\ref{sss:uniformb}), we cannot confirm this across multiple lines of sight (see \S\ref{sss:bvalues}).  Furthermore, while we find CH and CH$^+$ to be strongly correlated with different $J$ levels of $\Hmol$ (see \S\ref{ss:molecules}), the correlation is weak for $J=2$ and $J=6-7$.  Even the strength of the correlations for the $J$ levels which are correlated do not guarantee that the velocity structures are close enough to ensure this method is the most accurate.  Finally, we note that our data set would not be uniform in this case, because we do not have complete CH and CH$^{+}$ data for all of our lines of sight.

Quantitatively, the major argument against using the measured velocity information from CH and/or CH$^{+}$ is that it rarely improves the $\chi^2$ compared to the uniform $b$-value method, never does so by more than 10-20\%, and in several cases results in a significantly worse $\chi^2$ (factors of a few, up to $\sim7$ in the case of HD 199579).  Similarly, the quantitative argument in favor of the uniform $b$-value method is that the $\chi^2$ value it produces is never worse than $\sim$30\% of the best $\chi^2$ of all methods we explore, with differences of only a few percent being typical (and it is the best method for some lines of sight).  Furthermore, because of its nature, we do not see any unlikely population inversions or dramatically inconsistent $b$-values.

\begin{figure}[t!]
\begin{center}
\epsscale{1.00}
\plotone{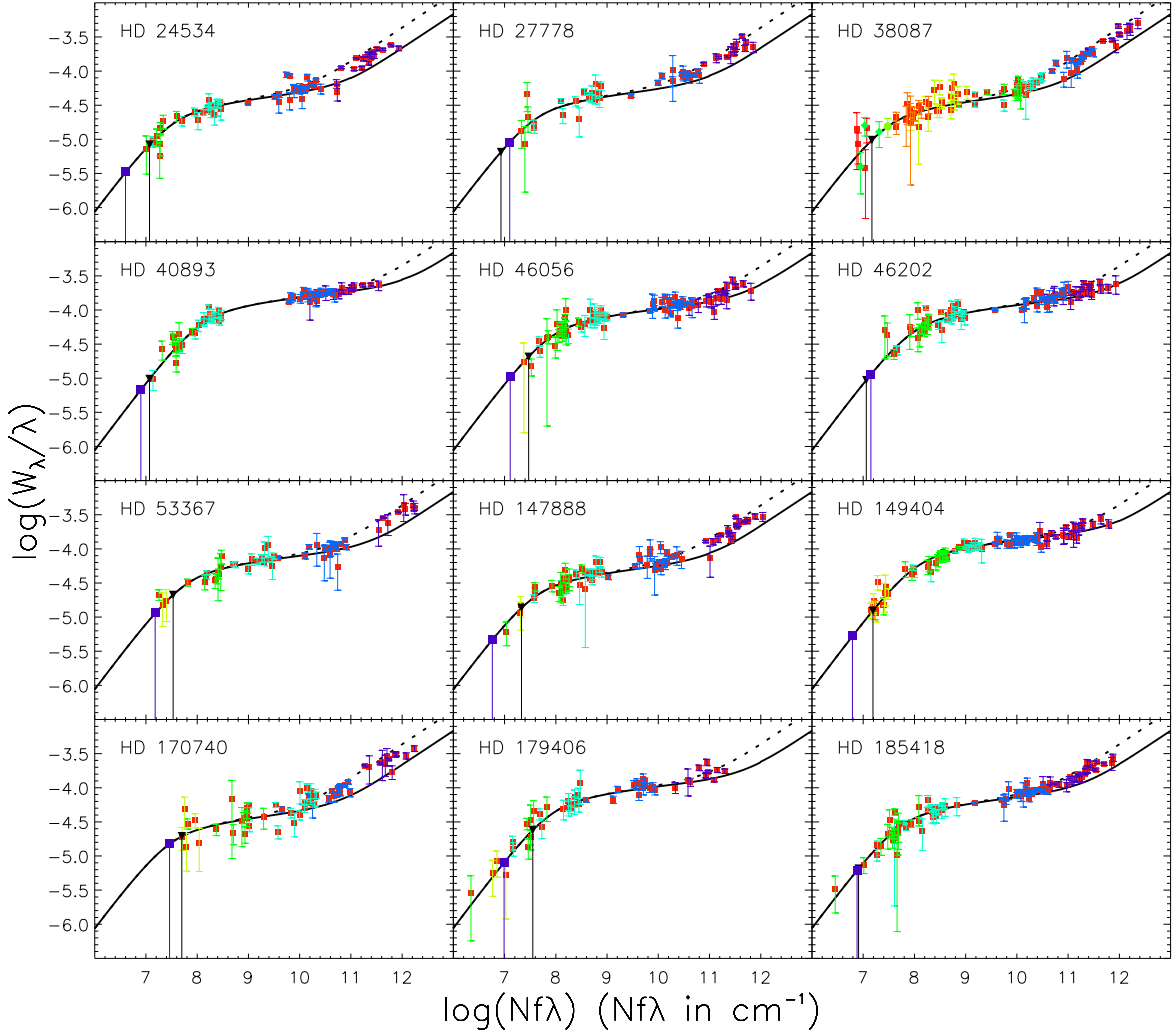}
\end{center}
\caption[HD 24534, HD 185418 Curve-of-Growth Results..]{Curve of growth fits for the HD 24534 through HD 185418 lines of sight, assuming a uniform $b$-value for all $J$ levels.  The error bars of the data points are color-coded by $J$ level.  The black lines are single component curves of growth with the $b$-value of the best fit solution.  The divergence of the lines to the right represent the range in values of the damping constant $\gamma$, which is different for each transition.  3-$\sigma$ limits of the first undetected $J$-level and the $J=0$, $\nu=1$ vibrational level are represented with black triangles and purple squares (with downward lines), respectively; detected vibrational lines are blue-green circles ($J=0$) and green diamonds ($J=1$), with matching colored error bars.  {\bf Purple:}---$J=2$.  {\bf Blue:}---$J=3$.  {\bf Light blue-green:}---$J=4$.  {\bf Green:}---$J=5$.  {\bf Yellow:}---$J=6$.  {\bf Orange:}---$J=7$.  {\bf Red:}---$J=8$.}
\label{fig:COGs1-12}
\end{figure}

\begin{figure}[t!]
\begin{center}
\epsscale{1.00}
\plotone{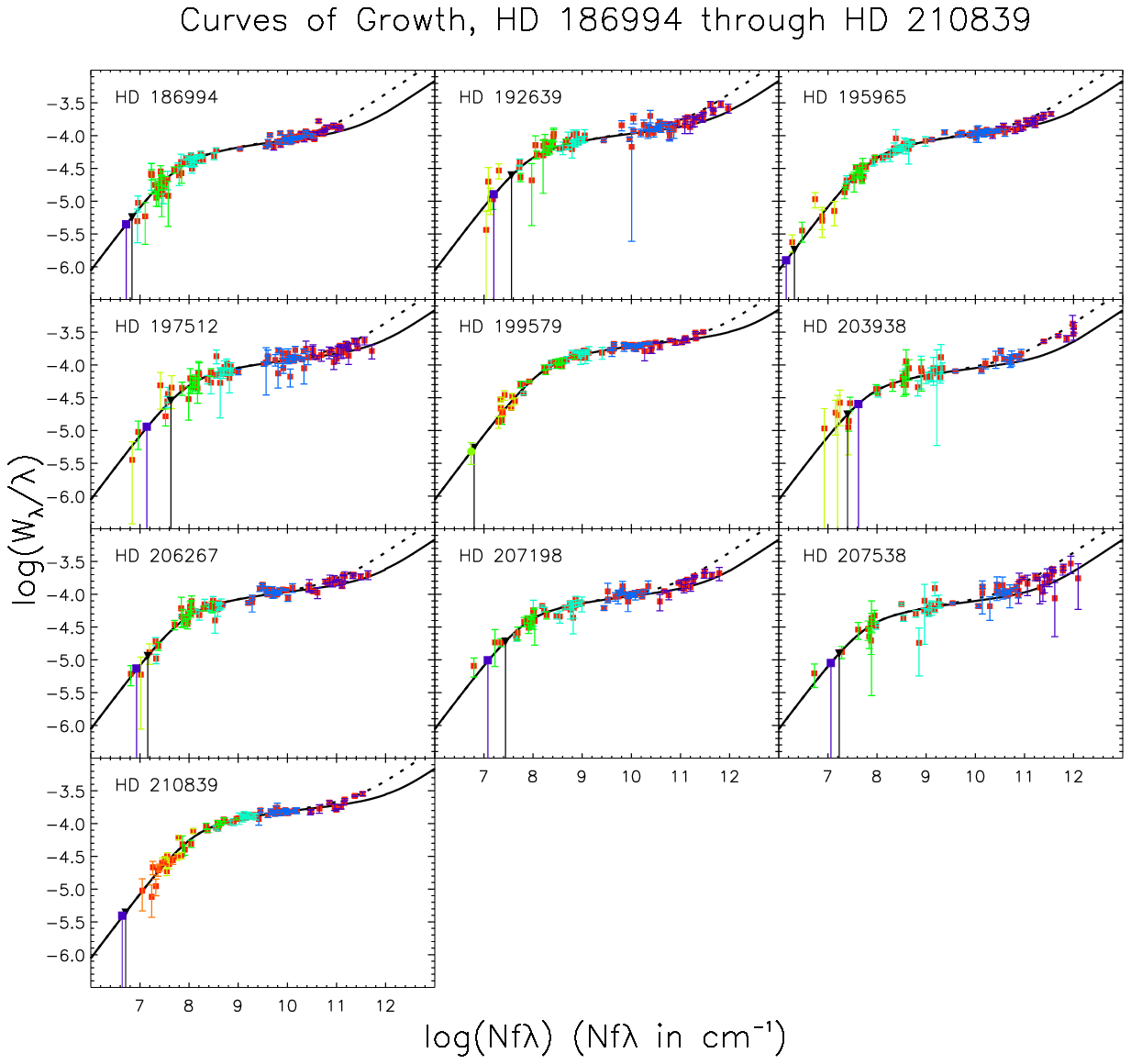}
\end{center}
\caption[HD 24534, HD 185418 Curve-of-Growth Results.]{Same as Fig.~\ref{fig:COGs1-12}, except for the HD 186994 through HD 210839 lines of sight.}
\label{fig:COGs13-22}
\end{figure}

Figs.~\ref{fig:COGs1-12}-\ref{fig:COGs13-22} show the curve of growth fits for all 22 of our lines of sight assuming a uniform $b$-value.  The fits to the various lines are color-coded by $J$ level.  Table \ref{coldensities} provides the column density results.  Note that the errors reported in this table, for computational simplicity, assume that the other column densities and $b$-value are held fixed, though technically the fitting of these column densities is dependent on the other column densities and their influence on the $b$-value.  Limits of the first undetected $J$ levels and vibrationally excited levels, and the few detections of vibrationally excited material, are shown in Tables \ref{limits}-\ref{vibrational}.

Though we adopt the uniform $b$-value method, comparison of the CH and CH$^+$ velocity structures with our adopted results are still instructive (we compare column densities in \S\ref{ss:molecules}).  For example, the five lines of sight where we derive the smallest $b$-value and also have CH and/or CH$^+$ velocity structures (HD 24534, HD 27778, HD 38087, HD 147888, and HD 170740) all have $b$-values of $\leq3.1\kmpers$.  These same lines of sight have only one or two detected CH/CH$^+$ components, separated by $\lesssim3\kmpers$.  Only HD 210839, where we solve for a $b$-value of $9.7\kmpers$, has a comparably small velocity extent in CH$^+$ ($\sim2\kmpers$), but the CH, K I, and Na I velocity structures show many more components with wider spacing in this line of sight.  Conversely, the largest $b$-value that we solve for, $12.5\kmpers$ in HD 199579, matches with the CH velocity profile with the widest spread of components ($\sim22\kmpers$); HD 149404, the third highest $b$-value (after HD 199579 and HD 210839) for lines of sight with CH/CH$^+$ data, is similar.  The remaining lines of sight with intermediate values for this study ($\sim4-8\kmpers$) are likewise intermediate in terms of the complexity and velocity extent of their CH/CH$^{+}$ profiles.

\begin{figure}[t!]
\begin{center}
\epsscale{1.00}
\plotone{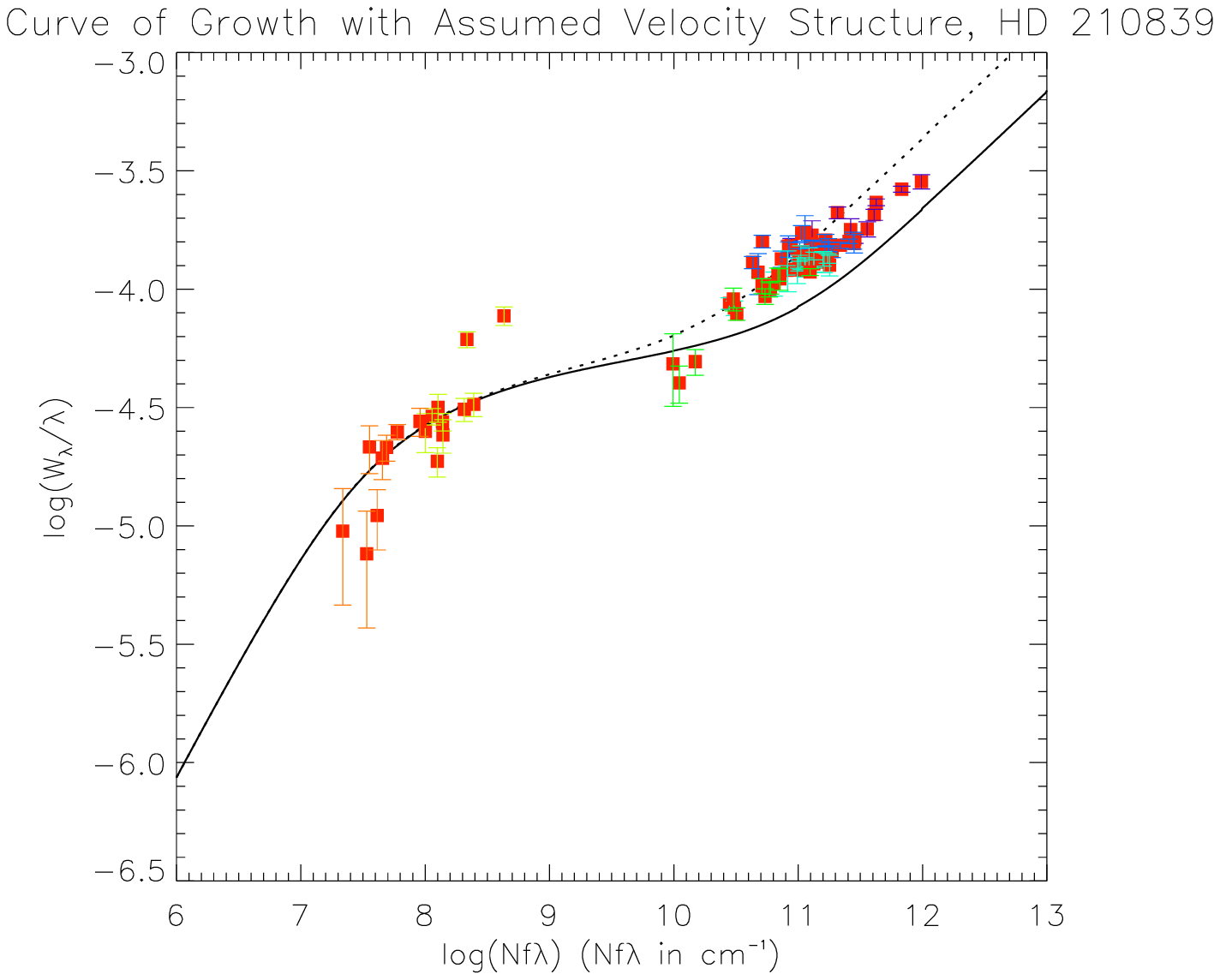}
\end{center}
\caption[HD 210839 Curve of Growth Fixed to Velocity Structure of CH.]{Same as Fig.~\ref{fig:COGs1-12}, except the specific case of HD 210839 when fixed to the velocity structure of CH.  The equivalent widths fit to the curve of growth in this line of sight much better with a much higher $b$-value than the velocity structure assumed here.  The results are similar whether CH or CH$^{+}$ data is used, and similar effects are observed in some other lines of sight, particularly HD 199579.}
\label{fig:210839example}
\end{figure}

However, note that this is a correlation of the velocity structures and not a strict correspondence.  In fact, as the structures become more complex, some of the fits to the observed CH/CH$^{+}$ structure are very poor.  Fig.~\ref{fig:210839example} shows the case of HD 210839 fit to the CH velocity structure.  In this case, there is a huge gap between the derived column densities of $J=5$ and $J=6$, and the fits for these column density levels are particularly poor.  Compare this to the last panel of Fig.~\ref{fig:COGs13-22}, where, while there is some scatter, the points follow a something close to single-component curve of growth (even though equivalent widths are of the measured profiles that appear to have at least components even in the {\it FUSE} data).  The situation is similar for the CH$^{+}$ velocity structure in this line of sight, and the CH/CH$^{+}$ structures in some other lines of sight, particularly HD 199579.

\subsection{Velocity Structure as a Function of $J$}
\label{ss:velocitystructure}
As we noted in \S\ref{s:intro}, the excitation of $\Hmol$ is so poorly understood that certain first-order types of questions about its velocity structure need to be answered.  For example, it is possible to construct a qualitatively simple model of a line of sight where, due to the excitation processes of $\Hmol$ and the physical structure of the cloud(s), different $J$ levels of $\Hmol$ essentially exhibit a distinct velocity structure.  For example, there might be two major clouds in a line of sight, one warm and one cold.  In this case, the former cloud might dominate the observed lines of higher-$J$ $\Hmol$ and the latter cloud the lower-$J$ lines.  Another simple possibility is that the higher-$J$ $\Hmol$ exists in a warm shell around the cold core containing the low-$J$ $\Hmol$.  In both of these examples, a possible (but not necessarily inevitable) observational consequence is that the simplest measures of velocity structure---$b$-value and velocity offset---may vary as a function of $J$.  Therefore, we conducted an examination of both of these quantities.

\subsubsection{$b$-values}
\label{sss:bvalues}
There are two basic ways to examine whether or not $b$-values vary as a function of $J$.  The first is to look at the apparent $b$-values of the Gaussian and Voigt fits.  The second is to examine the $b$-values derived through the curve of growth fits.

In examining the apparent $b$-values of the fits, we use only the 17 lines of sight for which profiles were fit with a single component.  We use a simple linear regression with $E_J/k$ as the abcissa, average $b$-value for a given $J$ level as the ordinate, and the standard deviation in the $b$-value for that $J$ level as the error.  We do this for each data channel of {\it FUSE} separately to provide independent checks on instrumental effects.  The slopes are nearly always negative and typically significant at between 1- to 3-$\sigma$ (there are two isolated exceptions of positive slopes in single data segments that are less than 1-$\sigma$ significant; in both cases, the other three data segments have negative slopes).  There are also a few cases of slopes that are more significant than 3-$\sigma$.  We calculate correlations with IDL's R\_CORRELATE function, which computes a Spearman's $\rho$ rank correlation coefficient.  A rank correlation is used in this case because we do not expect a particular functional form for any dependence.  Rather, we want to search to see if a roughly monotonic relationship exists---e.g.~$b$-value increasing as a function of $J$.  We also compute the probability of deviation from the null hypothesis, which gives similar results---nearly every data segment shows that an anticorrelation exists, typically significant at the 1- or 2-$\sigma$ level.

\begin{figure}[t!]
\begin{center}
\epsscale{1.00}
\plotone{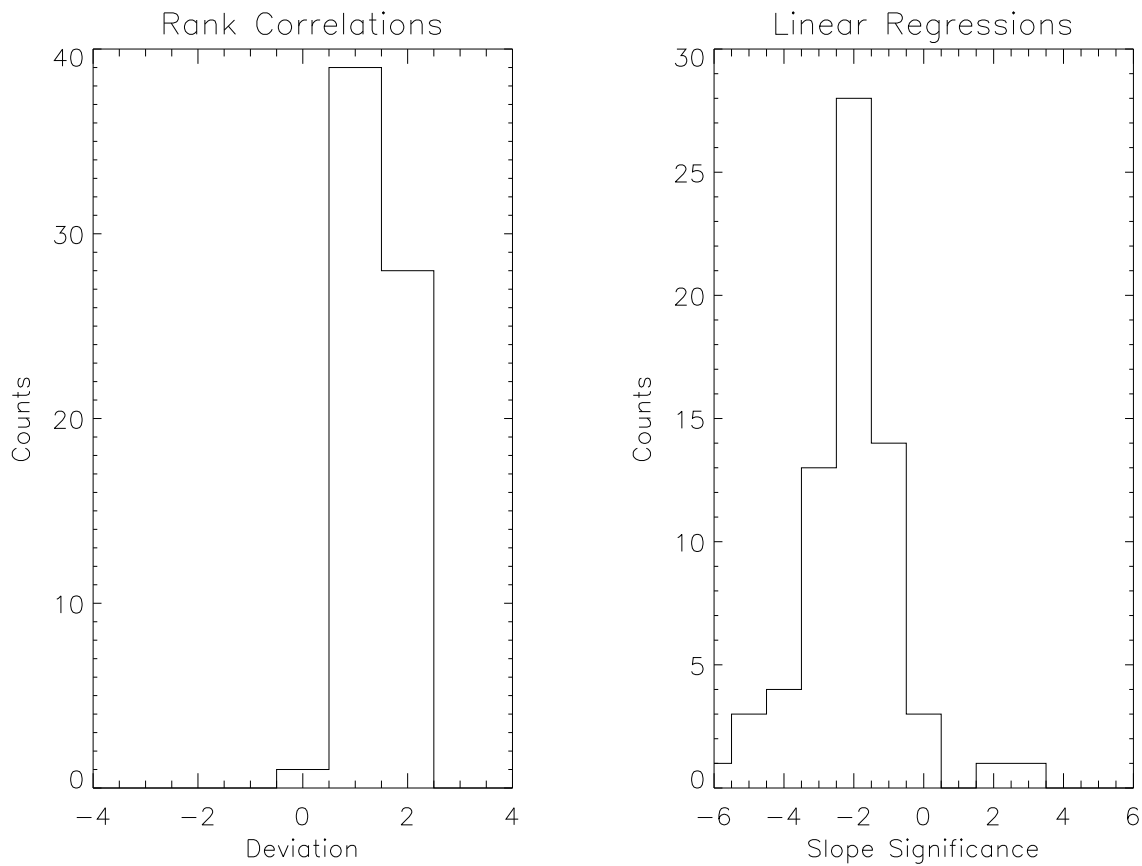}
\end{center}
\caption[Distribution of Correlation, $b$-values from Profile Fits.]{{\bf Left:}---The distribution of the calculated rank correlation coefficients of $b$-values vs.~$E_J/k$.  The abscissa is the significance of the rank correlation coefficients converted into a Gaussian cutoff value.  {\bf Right:}---The distribution of slope significance (slope divided by slope error) for the same data set.  Note that positive Gaussian cutoff values correspond to negative correlations/slopes.  Also note that there are 68 data points---17 lines of sight with four data segments each.  Reasons that there are apparent anticorrelations are discussed in the text.}
\label{fig:bcorr}
\end{figure}

However, we should note that when we analyze the unresolved apparent $b$-values of the fits, they are obscured by the broad resolution of {\it FUSE} ($\sim15\kmpers$).  This complicates the matter in two ways.  First, when the fits are Gaussian, the apparent width increases as a function of column density due to natural broadening, while still appearing very nearly Gaussian in the {\it FUSE} data.  Second, if there are weak components separated in velocity space, as these components become stronger with increasing column density the apparent $b$-value will also increase---that is, these components are lost in the noise for weak lines but blend with the dominant component(s) for strong lines, an effect that was noted in early studies of the ISM by \citet{RoutlySpitzer}.  Because the lower $J$ states have larger column densities, both of these effects could result in the negative slopes that we see---which say that $b$-value is decreasing as a function of $E_J/k$.  In addition, the {\it FUSE} resolution varies as a function of wavelength within individual detector segments, which introduces an additional level of uncertainty.  Together these effects significantly reduce our ability to use this method to see any real cases of $b$-values changing as a function of $J$, because at the {\it FUSE} resolution we cannot adequately account for these effects.  Fig.~\ref{fig:bcorr} shows histograms of both correlation coefficients and linear regressions, showing the systematic offset due to these effects; it can be seen in these histograms that additional deviations are likely no more than statistical variations.

Looking at the $b$-values derived from the curve of growth fits suffers from similar problems.  The apparent width due to natural broadening is not an issue (because the width is not being directly measured), but the possibility of an apparent increase in $b$-value with increasing column density (and therefore lower $J$ value) due to the weak component effect noted by \citet{RoutlySpitzer} still is.  In addition, the systematic uncertainties of assuming a single-component curve of growth come into play.

\begin{figure}[t!]
\begin{center}
\epsscale{1.00}
\plotone{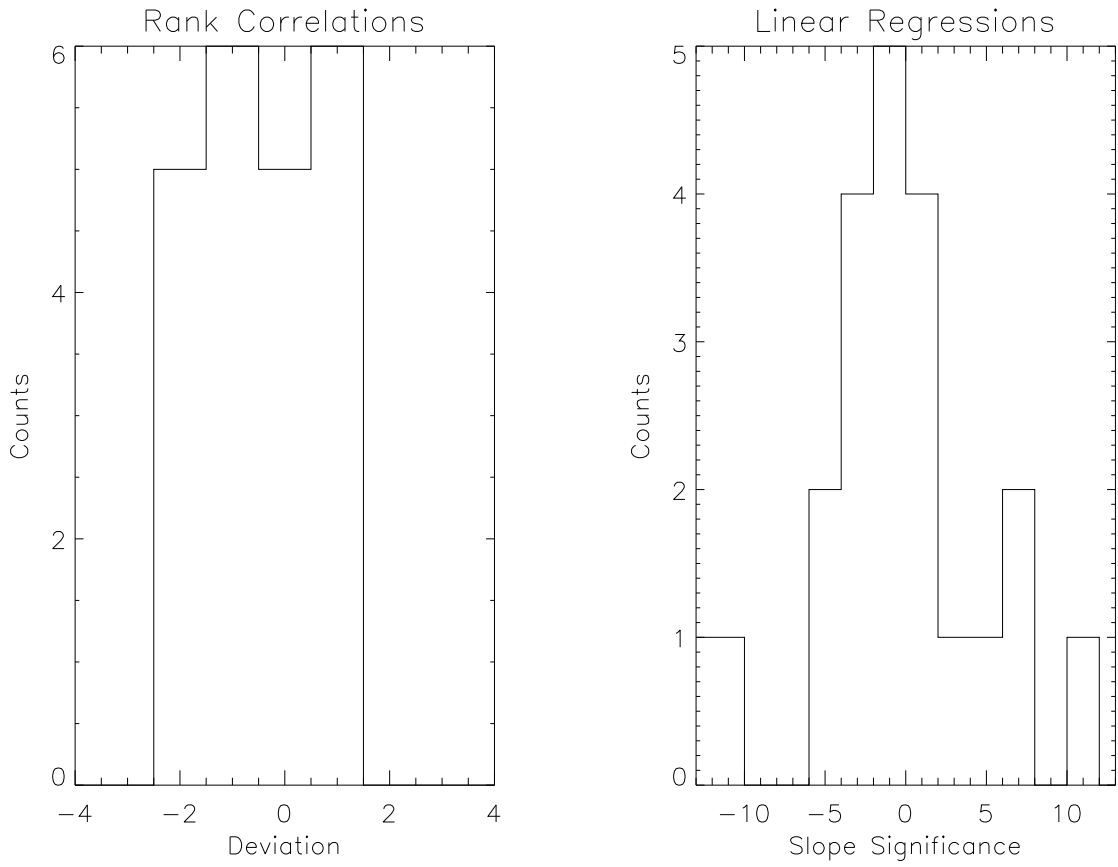}
\end{center}
\caption[Distribution of Correlations, $b$-values from Curve of Growth Method.]{Same as Fig.~\ref{fig:bcorr}, except using $b$-values from curve-of-growth method.  For this reason, there are only 22 data points.}
\label{fig:bcogcorr}
\end{figure}

Again using IDL's R\_CORRELATE function, we find that 12 of 22 lines of sight show correlations are less than 1-$\sigma$ significant; the other 10 are less than 2-$\sigma$ significant.  Exactly half of each level are correlations; the others are anticorrelations.  From this, we might conclude that any deviations are statistically distributed fluctuations.  When we use linear fits ($b$-value vs.~$E_J/k$), there are several (8) lines of sight that have greater than 3-$\sigma$ significant slopes.  However, these are mixed both positively and negatively.  The reason for the high confidence in the slopes may come from an underestimation in errors in the $b$-values.  The mixed signs of the slopes and correlations mean that we do not see a clear universal dependence of $b$-value on $J$ (in particular, we might expect an increase of $b$ with $J$ due to formation processes).  This is particularly true given the potential systematic errors that we have discussed, though at the same time weakens the confidence with which we can state the null result.  Fig.~\ref{fig:bcogcorr} shows histograms of the correlation and regression methods, showing that the deviations may be explained as roughly statistical variations.

\subsubsection{Velocity Offsets}
\label{sss:velocityoffsets}
We also examined velocity offsets as a function of $J$ in each line of sight.  We take an average of the velocity offsets of all the lines of a given $J$ level in each line of sight and in each {\it FUSE} channel.  Keeping the {\it FUSE} channels separate is more important for velocity offsets than for $b$-values because there are systematic offsets between the channels which our initial data reduction does not correct (it is relatively unimportant other than for this question).  Again, it also allows us to perform a check on the significance of any possible trends---a slight trend may be seen in one data channel but not in the others.  As in \S\ref{sss:bvalues}, we only perform this analysis for the 17 lines of sight where we used single-component fits to measure the velocity offset.

One of us (BLR) measured velocity offsets for the $J=0-1$ lines from previous data as part of the FTCS, but there are problems with trying to use these data in our comparisons.  First, some of these fits were made with very early versions of the CALFUSE pipeline reduction that may not be consistent with ours, which has the potential to be very significant for measuring velocity offsets.  Second, these previous fits included the $J=2$ lines, meaning that the velocity measurements of the $J=0,1$ lines are not completely independent.  Therefore, we instead measure the HD lines that are in the {\it FUSE} wavelength range.  This is in part justified by the very strong correlation that we find between HD and $\Hmol$ (see \S\ref{ss:molecules}).  We use the same HD lines discussed in the recent survey of HD by \citet{Snow2008}, except that we omit the weakest line at 1105.834 \AA{} because it is generally too weak to detect and when it is detected it suffers from confusion with neutral carbon lines.  Because we measure the $J=0$ lines of HD, we use these as a proxy for $\Hmol$ in the $J=0$ state and omit $J=1$ from this portion of the analysis.

When we examine only $J \ge 2$, there are few correlations to be seen.  There are no cases where as many as three of the four data segments for a given line of sight have greater than 1-$\sigma$ significance (judging by slopes); occasionally 1- or 2-$\sigma$ significant correlations appear in one or two individual segments, but are not supported by the remaining two or three data segments.  Plotting a histogram of the 68 correlation levels and 68 slope significances\footnote{Four data segments for each of the 17 lines of sight that use only single-component fits to the absorption lines.} indicates that any exceptions are easily explained as statistical variations.

\begin{figure}[t!]
\begin{center}
\epsscale{1.00}
\plotone{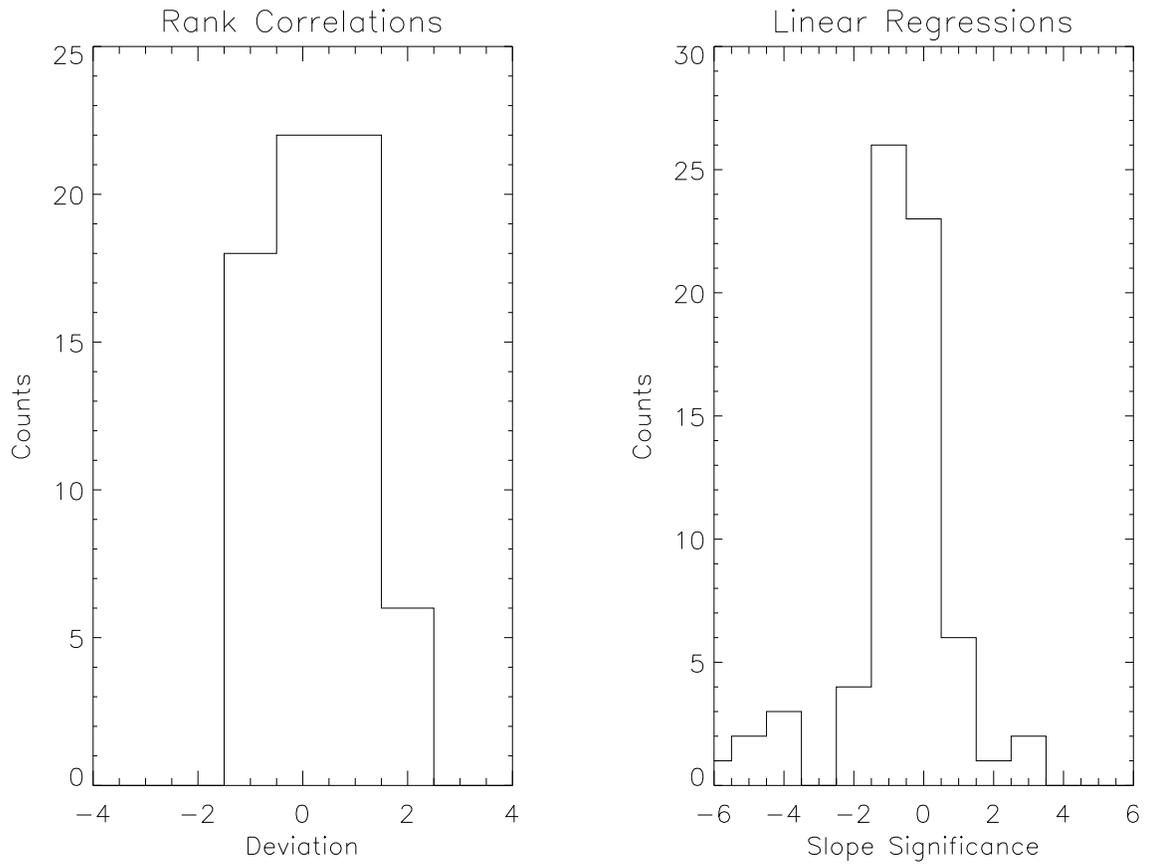}
\end{center}
\caption[Distribution of Correlations, Velocity Offsets.]{Same as Fig.~\ref{fig:bcorr}, except for velocity offsets.}
\label{fig:vcorr}
\end{figure}

However, when the HD lines are included as a proxy for $J=0$, the situation becomes more complicated.  The rank correlation coefficients are centered around a positive 1-$\sigma$ deviation, and the slope significances are centered around negative 1.5-$\sigma$.\footnote{The correlation coefficients are translated into probability deviations; positive deviations correspond to negative correlations/slopes.}  Though this superficially implies that there is a discrepancy between the location of the coldest $\Hmol$ (where the HD is) and the higher-$J$ $\Hmol$, there are issues to consider.  First, if there is such an effect, we would not expect it to be manifested as a shift toward either positive or negative correlations that holds across all lines of sight, we would instead expect that there are more outliers that can be explained statistically when assuming a null hypothesis.  The width of the correlation/slope significance histograms stays constant in our case.  Next, we measure a relative paucity of HD lines compared to $\Hmol$ lines---we only search for six lines, and in some cases only one or two lines are measured due to poor S/N.  Therefore, it is possible that there are systematic errors in the measured velocities of the HD lines, either in our reference wavelengths or due to the instrument.  We must consider that the {\it FUSE} resolution is $15\kmpers$ (FWHM) and that there may be comparable uncertainties in the {\it FUSE} wavelength solution.  While our measurements of many lines are intended to reduce these errors through improved statistics, they are still significant.  The average slope measured in the HD-inclusive case, integrated between the $J=0$ HD and the maximum $J$-level of $\Hmol$ detected for each line of sight, is $10\kmpers$.  Finally, extending the correlation down to $J=2$ should be sufficient to detect differences between the coldest and warmer material, as $N(2)$ is consistent to a degree with the excitation temperature implied by $N(0)$ and $N(1)$, further supporting the possibility that the HD lines introduce systematic errors.  Fig.~\ref{fig:vcorr} shows histograms of the correlation coefficients and regression significances.


Ultimately we conclude that we do not observe any definite cases of velocity offset varying with $J$, as the HD-inclusive velocity results, while intriguing, are more likely due to systematic errors.  It is very important to note that the imprecision of the {\it FUSE} wavelength solution and its moderate resolution prevent us from stating a null result more strongly.  However, our results are still suggestive of the conclusion that the higher- and lower-$J$ $\Hmol$ are not dominated (in an observational sense) by physically distinct regions.

\section{DISCUSSION}
\label{s:discussion}
We now turn our attention to other important issues regarding the interpretation of our measurements:  how does higher-$J$ $\Hmol$ correlate with other molecules, what is the temperature of these clouds (and what is the implication for formation and destruction processes), and can these results be modeled theoretically?

\subsection{Correlations with Other Atomic and Molecular Species}
\label{ss:molecules}
CH is known to be correlated with total $\Hmol$ (dominated by $J=0-1$; see the FTCS and references therein).  In lines of sight with less extinction, CH$^+$ is known to correlate with higher-$J$ $\Hmol$ \citep{LambertDanks}.  What has not been explored in detail at higher extinction is how higher-$J$ correlates with CH, CH$^+$, or other atoms and molecules.  We have taken available column densities for several of these atomic and molecular species and attempted to find existing correlations.

Our original working list of atoms and molecules\footnote{These data are primarily taken from a database of column densities by D.~E.~Welty available at http://www.astro.uiuc.edu/$\sim$dwelty/coldens.html.  Many of these data are published in \citet{Welty2001, Welty2003, Sonnentrucker2007}.  Exceptions are that, where available, Fe II is taken from \citet{JensenFeII} and HD is from the single-component curve of growth results from \citet{Ross2008}.} included C I, S I, K I, Li I, Ca I, Ca II, Na I, Fe I, Fe II, Ti III, C$_2$, C$_3$, $^{12}$CO, $^{13}$CO, CN, HD, CH, and CH$^+$.  Obviously, only some of these are expected to be highly correlated with $\Hmol$ regions, while others may give an indication of the radiation levels.  Notably, the database we are drawing from does not overlap perfectly with ours, meaning that in some cases we are attempting to correlate a very small number of points.  Though we calculate the probably that correlations are not null based on the number of available data points, any correlation with a significantly small number of data points should be considered with healthy skepticism.  We have therefore chosen to only include in our calculations those atoms and molecules for which there exist at least 10 lines of sight overlapping with our study (and therefore $\geq10$ common data points for all correlations involving $J\leq5$; correlations involving $J\geq6$ may have fewer points).  This revised list is Li I, K I, Ca I, Fe I, Fe II, HD, CH, CH$^{+}$, $^{12}$CO, C$_{2}$, and CN.

Tables \ref{corrtablecols}-\ref{corrtableJ} show correlations between these species and various measures of the $\Hmol$ column density and population distribution.  All the tables show the Pearson correlation coefficient $r$ (positive is a correlation, negative is an anticorrelation); the two-sided confidence level that the correlation is nonzero for a $t$-distribution of the quantity $r(n-2)^{1/2}(1-r^2)^{-1/2}$ for $n-2$ degrees of freedom, where $n$ is the number of data points; and finally, $n$ is also listed for each correlation calculation.


Tables \ref{corrtablecols}-\ref{corrtableratios} have the same columns but different rows.  The first column is total $\Hmol$ column density, and the rest are several different ratios of the column densities of different $J$ levels.  The various ratios probe different qualities of the lines of sight.  $N(2)/N(0)$ and $N(3)/N(1)$ are indicators of the gas density; greater ratios imply lower density, because $J=2-3$ are collisionally deexcited \citep{ShullBeckwith}.  Similarly, $N(4)/N(0)$ and $N(5)/N(1)$ are indicators of the radiation field, because $J=4-5$ column densities are set largely by the radiation field, so larger ratios imply more radiation \citep[again, see][]{ShullBeckwith}.  The other three ratios that we include in this table are $N(4)/N(2)$, $N(5)/N(3)$, and $N(6)/N(2)$.  These measures in essence merge the more commonly used ``radiation" and ``density" ratios that we just discussed to give combined measures of the two quantities; the last quantity, $N(6)/N(2)$ does so by looking at $N(6)$ to examine the top end of the radiation, but is limited by the number of $J=6$ detections.  Table \ref{corrtablecols} shows the correlations between the parameters discussed in the previous paragraph with the column densities of the atomic and molecular species.  Table \ref{corrtableratios} explores three potentially interesting ratios of the atomic and molecular species:  CH$^+$/CH, CH$^+$/CN, and Fe II/Fe I.

In Table \ref{corrtablecols}, the correlations for Li I, Ca I, Fe I, Fe II, C$_2$, and CN are less than $<95$\% significant,\footnote{That is, the probability that the calculated correlation coefficients are not from a normally distributed population is $<95$\%.  In the discussion that follows, 95\% is the nominal criterion for ``significance," although some exceptions are discussed.} with only the correlations between total $\Hmol$ and C$_2$ and CN $>90$\%.  K I shows a significant correlation with total $\Hmol$, and significant anticorrelations with the density and radiation ratios.  All of these are unsurprising, as we expect K I, with its lower ionization potential, to only exist were it is relatively dense and shielded.  HD shows a similar pattern, although the correlations for the density and radiation ratios do not meet the criterion for significance as they do for K I [although $N(5)/N(1)$ is close at 94.7\%].

CH, as already known, is very strongly correlated with the total $\Hmol$, but it is almost as strongly anticorrelated with the combined ratio of $N(6)/N(2)$.  Anticorrelations are also observed for the density and radiation ratios, although only $N(3)/N(1)$ meets our significance criterion.  For CH$^+$, significant correlations are observed with the radiation ratios, and two of the combined ratios are significant or nearly so [94.5\% for $N(4)/N(2)$, 97.7\% for $N(5)/N(1)$].  Combined with the fact that the correlations with the density ratios are insignificant, this stresses the possible importance of radiation in CH$^+$; at the very least, it confirms the connection between CH$^+$ and $J=4-5$, including potentially between their formation processes.

Finally, $^{12}$CO has a correlation pattern that is basically the opposite of CH$^+$---significant anitcorrelations for $N(4)/N(0)$ and $N(4)/N(2)$ and nearly significant anitcorrelations for $N(5)/N(1)$ and $N(5)/N(3)$ (just over $90$\% but not meeting our significance criterion).  As photodissociation is the dominant destroyer of CO, these anticorrelations are not surprising to the extent that the radiation ratios and combined ratios do actually correspond radiation levels.

Moving on to Table \ref{corrtableratios}, of the three ratios shown in the rows, only CH$^+$/CH shows significant results.  The correlations with the radiation ratios are very strong ($\geq99.5$\%), as would be expected based on the correlations with CH and CH$^+$ separately.  The other correlations also follow according to expectation, although at reduced significance---the combination ratio $N(4)/(2)$ is still significant at 99.0\%, while the combination ratio $N(5)/N(3)$ and the two density ratios are $>90$\% but do not meet our significance criterion.  Once again, we note the strong connection to $J=4-5$ and, implicitly, to the radiation levels.

Table \ref{corrtableJ} shows column density correlations as a function of $J$-level.  In this table, the correlation coefficient is modified to account for the total hydrogen column density [$\NHtot = \NHI + 2\NHmol$] of these lines of sight.  That is, we adopt the formalism used by \citet{Jenkins1986} in order to make sure that correlations are not the spurious result of total column density---if this step is not taken, all of the correlations are very, very strong.  We also note that in doing this, there may be some effect of inhomogeneity (and therefore possible systematic errors) of the atomic hydrogen sources.  The references for all of the atomic hydrogen values that we use, along with some discussion of concerns of inhomogeneity, can be found in \citet{JensenFeII}.

Many of the correlation coefficients of Table \ref{corrtableJ} are statistically insignificant, but there are exceptions.  All the $J$ levels are anticorrelated with Fe I, with the confidence levels meeting our selection criterion for $J=2-5$.  This is most likely explained as an effect of radiation ionizing the Fe I to Fe II.  K I follows the correlation patterns seen in Table \ref{corrtablecols} (correlated for $J=0-1$, anticorrelated for $J\geq2$).  Only $J=2$ and $J=6$ meet our significance criterion; however, all other levels except $J=7$ are $\gtrsim90$\%.

The molecules are likewise consistent with what we would expect.  HD is correlated with $J=0-1$ and anticorrelated with $J=5-6$; only $J=0$ meets our significance criterion, but the other three are $>90$\%.  CH is strongly correlated with $J=0-1$ (meeting our criterion) and weakly anticorrelated with $J=6-7$ (failing to meet the criterion in part because of fewer data points; the $J=7$ coefficient is actually quite large).  CH$^+$ is correlated with $J=3-5$ (at 94.9\%, 98.9\%, and 99.9\%, respectively); none of the other correlations or anticorrelations meet the significance criterion.  And lastly, the correlation coefficients roughly increase as a function of $J$ for $^{12}$CO---correlated for $J=0$ and anticorrelated for $J\geq4$.  Although only $J=5$ meets the significance criterion, $J=0$, $J=4$, and $J=6$ are all $\gtrsim90$\% ($J=6$ is less significant than $J=5$ despite having a larger correlation coefficient due to fewer data points).

\subsection{Excitation Temperature}
\label{ss:Texc}
A well-established observation of $\Hmol$ is that the column densities of the various $J$ levels are rarely able to be fit by a single excitation temperature \citep[e.g.~see][]{SpitzerCochran}, although exceptions are found in lower-column density, higher-temperature lines of sight \citep[e.g.~][]{MortonDinerstein}.  Excitation temperature is defined by
\begin{equation}\label{eq:texc}
\frac{N_u}{N_l} = \frac{g_u}{g_l}e^{-\Delta E / kT_{exc}}
\end{equation}
where $N$ is the column density, $g$ is the Gaunt factor (statistical weight), the $u$ and $l$ subscripts are the upper and lower state values, $\Delta E$ is the difference in the energy levels (upper minus lower), and $T_{exc}$ is the excitation temperature.  If the logarithms of the statistically-weighted column densities is plotted against $\Delta E/k$, then this relationship should be linear for constant $T_{exc}$.  However, typical plots show that this relationship is never linear from $J=0$ extending to $J=5$ and beyond.  Rather, most lines of sight are best fit with a sum of populations at different temperatures.  Although the best fit may require more than two temperatures, it is typical to use one temperature from $J=0$ to $J=2$ or 3, and another for higher $J$-levels.  Because of number of our data points (typically only 6 or 7 $J$-levels) and the number of free parameters required (one temperature and one normalization for each population assumed), we will limit our exploration to only two temperatures.

Our first step is to determine the most appropriate break point for this fit.  Defining $J_{break}$ as the $J$ level where we assume $J < J_{break}$ is fit with one temperature and $J \geq J_{break}$ is fit with another, we test $J_{break}=2$, 3, and 4.  In every line of sight, $J_{break}=3$ produces the lowest global reduced $\chi^2$ ($\equiv \chi_{\nu}^2$)---although $\chi_{\nu}^2$ is rarely below 1, revealing that either errors have been underestimated or this simple model is inadequate.

\begin{figure}[t!]
\begin{center}
\epsscale{1.00}
\plotone{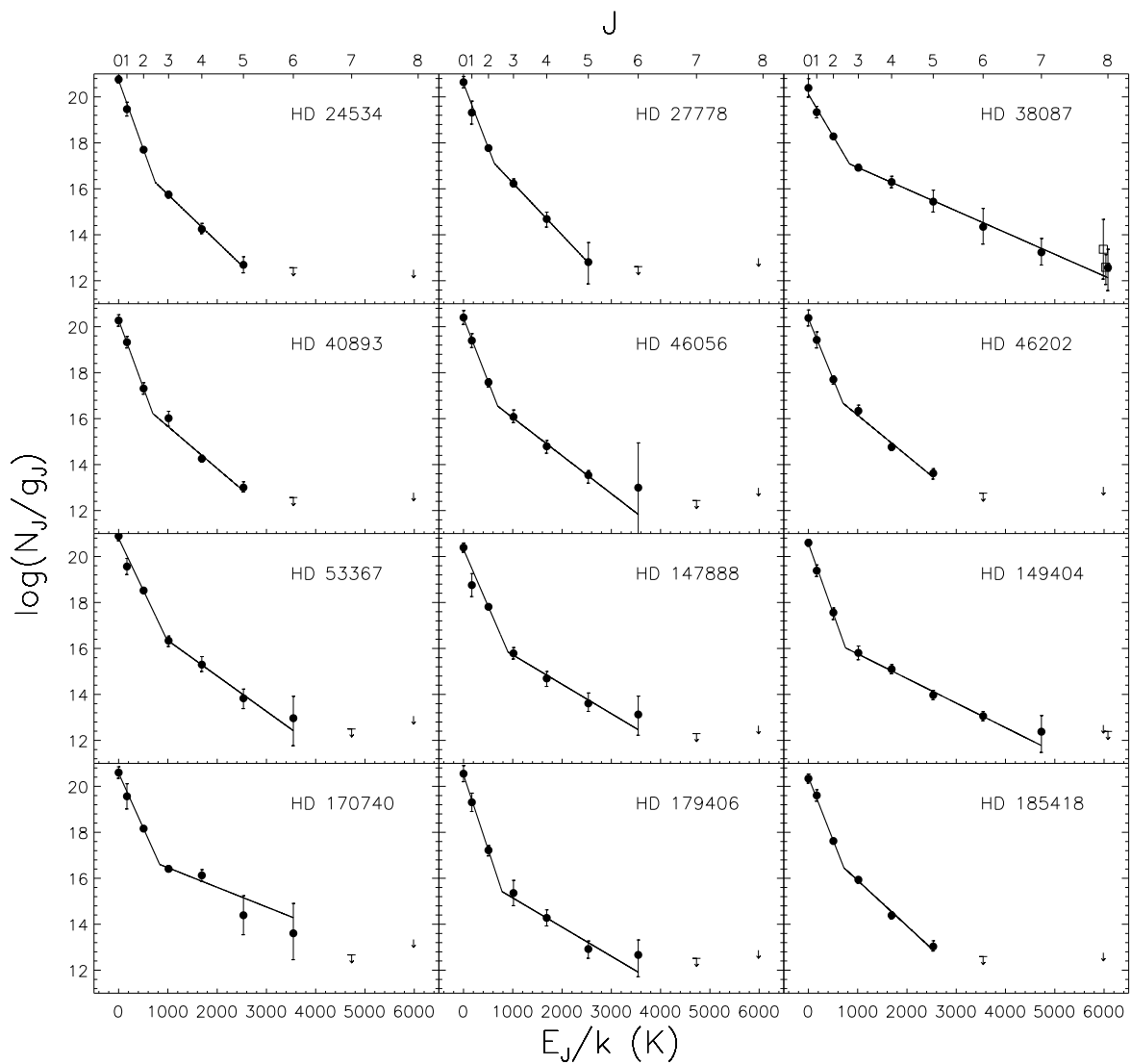}
\end{center}
\caption[Excitation Temperatures, HD 24534, HD 185418.]{Fits of two excitation temperatures to statistically weighted column densities for the HD 24534 through HD 185418 lines of sight.  Errors are exaggerated on the plot's logarithmic scale by a factor of 5 for clarity.  Filled circles are the main $J=0-8$, $\nu=0$ states.  The 3-$\sigma$ upper limit on the first undetected $J$-level is a downward arrow with a horizontal bar.  3-$\sigma$ upper limits on the $J=0$, $\nu=1$ state are shown with downward arrows without a bar.  Detections of vibrationally excited material are shown with open squares, only applicable for HD 38087 on this plot and HD 199579 in Fig.~\ref{fig:texc13-22}.  See text for further discussion.}
\label{fig:texc1-12}
\end{figure}

\begin{figure}[t!]
\begin{center}
\epsscale{1.00}
\plotone{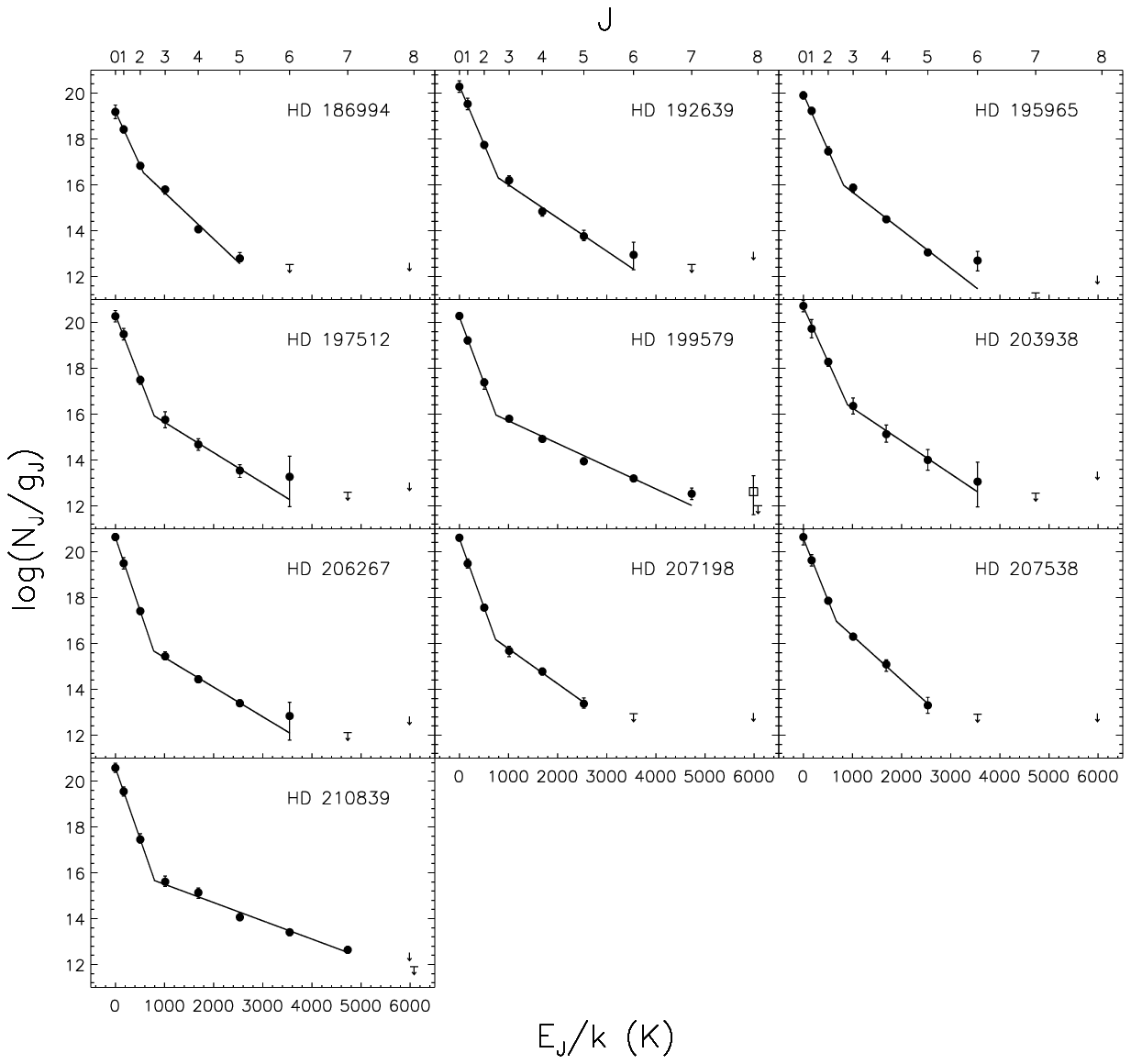}
\end{center}
\caption[Excitation Temperatures, HD 186994, HD 210839.]{Same as Fig.~\ref{fig:texc1-12} except for the HD 186994 through HD 210839 lines of sight.}
\label{fig:texc13-22}
\end{figure}

Figs.~\ref{fig:texc1-12}-\ref{fig:texc13-22} and Table \ref{texctable} show our fits assuming $J_{break}=3$.  For comparison, $T_{01}$, derived solely from the $J=0-1$ column densities, is also shown.  It can be seen that $T_{02}$ and $T_{01}$ are not entirely consistent---only five lines of sight agree within their 1-$\sigma$ errors.  However, the discrepancies are still instructive---in the five worst cases, the ones which disagree by more than 3-$\sigma$, $T_{02} > T_{01}$.  This is what would be expected if the processes responsible for populating the $J \geq 3$ levels of $\Hmol$ also have an impact on $J=2$.

We also note that for 19 of our 22 fits, the $N(3)$ is greater than the fit; in 14 of those cases, the discrepancy is greater than 1-$\sigma$.  Combined with what was discussed above about the difference in $T_{01}$ vs.~$T_{02}$, this is evidence that $N(2)$ and $N(3)$ are best fit by a third temperature.  However, it is difficult to put quantitative certainty on this because the number of data points is too small, as discussed above.  We also note that the relationship of $N(4)$ and $N(5)$ to the fit do not seem to indicate an ortho-para effect; both $N(4)$ and $N(5)$ tend to be underestimated by the fit greater than half of the time, although several points are within the errors.

Judging by the plots, it is also not infrequent that the column density of the highest $J$ level commonly detected---$J=6$---is underestimated by the fits.  There is certainly the possibility that $\Hmol$ in the $J \geq 6$ states is better fit by yet an another temperature, but again by adding too many temperatures we lose our ability to ascertain the quality of the fit.  Notably, $N(8)$ is also underestimated by the fit for HD 38087, and $N(7)$ is underestimated by the fits for HD 149404, HD 199579, and HD 210839 (the underestimation in smallest in the last case)---all cases of the column density of the highest detected $J$ level being underestimated.  A second possibility is that our column densities for the limiting cases are overestimated due to random noise enhancements of the relatively few detected lines (though we note we are confident that the detections are real due to consistency with velocity offsets).

The weighted average value of $T_{02}$ is 77 K, which is higher than the average $T_{01}$ determined in the FTCS---reasons for this discrepancy were discussed above.  The weighted average of $T_{3+}$ is 278 K, which is consistent with previous surveys \citep{SpitzerCochran, Spitzer1974}.


\subsection{Meudon PDR Code Modeling}
\label{ss:PDRcode}
As noted in \S\ref{s:intro}, \citet{Nehme2008} used the Meudon PDR code \citep[most recently described by][]{LePetit2006} to model the HD 102065 line of sight.  They were able to replicate certain observational properties of this line of sight, but not the higher-$J$ $\Hmol$ or CH$^+$ column densities.  Increasing the radiation improved the fit to these excited column densities, but resulted in a worse overall fit to all of the observational quantities.


We also used the Meudon PDR code to model two of our lines of sight.  We take an approach similar to \citet{Nehme2008}:  we fix all the possible inputs ($A_V$, $R_V$, total hydrogen column density, etc.) and allow only the density and radiation field to vary.  Our grid is very coarse, but covers a wide range in each variable.  The local density $\nHvol$ is allowed to vary from $20-10^{4}\percc$, and values of the radiation field are allowed to vary from $0.14-140$ G (where G is the average Galactic value assumed by the code).  The spacing is logarithmic, by factors of $\sim1.5$ ($\sim0.15\dex$ on the logarithmic scale).

We selected two lines of sight, HD 46056 and HD 147888, as test cases.  Neither line of sight necessarily represents a ``typical" line of sight in our target list.  It would be incorrect to say that such a line of sight exists in our sample, as many of the lines of sight were chosen for the FTCS due to their interesting properties.  HD 46056 is in the open cluster NGC 2244 and has a slightly below-average value of $\rv$ and the subsequently expected higher far-UV extinction.  In addition, the 2175 \AA{} extinction feature is weaker and narrower than in other lines of sight, qualities often associated with steep far-UV extinction.  HD 147888 is notable as a sight line that passes through the $\rho$ Oph complex,\footnote{HD 147888 is also known as $\rho$ Oph D.} and at $\sim130$ pc has one of the shorter line of sight pathlengths in the FTCS.  HD 147888 has a very large value of $\rv$ ($> 4$), and therefore less far-UV extinction.  The fixed parameters for these lines of sight are listed in Table \ref{modelinput}.

We perform the $\chi^2$ calculation using logarithmic column densities and errors of both $\Hmol$ and the other molecules.  This is done for two reasons.  First, it weights the measured column densities more evenly, in better proportion to the true uncertainties.  Second, when the calculation is performed using linear column densities, the best fit corresponds to a point at the density edge of our grid (and near the radiation edge) in both lines of sight.  This was true for an earlier smaller and even more coarse grid that we calculated, but expanding the grid and increasing its resolution did not resolve the issue.  However, a local minimum in the $\chi^2$ contour plots was found elsewhere in the grid for both lines of sight.  These local minima are similar to the minima obtained when considering logarithmic column densities.



The solutions for both lines of sight, reported in Table \ref{modeloutput} are very high-density, high-radiation solutions.  This is quite dissimilar to the results found by \citet{Nehme2008} for HD 102065.  We note that our solutions to both lines of sight are poor overall, with $\chi_{\nu}^{2} \gg 1$, although some of the $\Hmol$ column densities are nearly within the errors for HD 147888.  In particular, the $J=0-1$ column densities, molecular fraction of hydrogen, and the ortho-to-parahydrogen ratio are very poorly fit.  Our best fit models also severely underpredict the CH$^+$ column densities, similar to the \citet{Nehme2008} models, and overpredict the CH column densities.

Furthermore, with regard to HD 147888, we note that \citet{Sonnentrucker2007} also made an estimation of the density for this line of sight from C$_2$ excitation that is significantly smaller ($215\percc$) than our solution.  This result assumed a typical interstellar radiation field, although lower K I and Na I abundances in this line of sight also imply an above average radiation field which our results do imply.

Both our results and the \citet{Nehme2008} results show the difficulty of applying the Meudon PDR code to the issue of $\Hmol$ excitation.  The quality of the \citet{Nehme2008} fit is much better than ours---their result for HD 102065 had a value of $\chi^2_{\nu}$ that was less than 1, while our results for the $\chi^2_{\nu}$ of the best fit are significantly greater than 1.

One important shortcoming of the Meudon PDR code, which is also true of many radiative transfer codes, is that such codes are usually steady-state models.  However, the ISM is quite dynamic and rarely in complete local thermodynamic equilibrium.  Specifically, shocks are not considered.  \citet{Nehme2008} concluded that shocks or other dissipation of turbulent thermal energy are the best explanation for the elevated levels of CH$^+$ and higher-$J$ $\Hmol$ relative to their best fit model's predictions.

This same concern applies to our results as well.  However, it is only one of several possible explanations for the discrepancy of our measurements with the results of our test models.  These include the following, listed roughly from greater to lesser probable importance:

\begin{enumerate}
\item The steady-state assumption of the Meudon PDR code, especially the lack of shocks
\item The assumption of a single, uniformly-dense cloud as opposed to multiple cloud components of varying density
\item Errors in the input measurements, e.g.~the column density of H I (particularly for HD 147888, which has an unusually high value of $\NHI$ compared to the other line of sight parameters and is uncertain due to the late spectral type of the star)
\item Possible systematic errors have not been quantitatively accounted for, which would at least reduce the $\chi^2$
\item Chemistry and dust parameters were not tweaked, but were left at default values---although the values are reasonable and consistent with Galactic averages
\item The coarseness of our grid makes any solution less than optimal
\end{enumerate}

We note that a different radiative transfer code by \citet{BTS2003} was able to produce much better results for the higher-$J$ $\Hmol$ using either high radiation fields or ``concatenated" models, that is, assuming a superposition of clouds as seen by the observer.  However, one major difference between the \citet{BTS2003} and the \citet{Nehme2008} is that the former requires the input of both a density and a temperature which are assumed to be uniform throughout the cloud, while the latter more realistically only assumes a uniform density (which still may be a poor representation) and calculates the temperature throughout the cloud.  The \citet{BTS2003} fits also did not consider the other observational data that is available that, in principle, any adequate model should be able to match.

It remains to be seen whether a concatenation in the manner of \citet{BTS2003} can be used in conjunction with a more complex code to consistently produce better results for a wide range of lines of sight.  \citet{Nehme2008} do attempt this for their line of sight, but find that it requires a second cloud of warm ($\sim250$ K), dense ($\nHvol \geq 10^4 \percc$) gas within 0.03 pc from the background star that has a ortho-to-para hydrogen ratio at formation near unity.  \citet{Nehme2008} note, in addition to the extreme nature of these conditions, that IRAS observations imply a distance of the cloud at least 0.12 pc from the star and an ortho-to-para formation ratio of $\sim1$ is theortetically and experimentally unfounded.  \citet{Nehme2008} therefore conclude that this scenario is not plausible, and favor the explanation of the dissipation of thermal energy to explain the higher-$J$ and CH$^+$ results.  We further note that given the qualitative similarity of our column density results for other lines of sight, this solution is even less plausible as a widespread explanation for the observations.

We note that through this work we have substantially increased the higher-$J$ $\Hmol$ measurements for lines of sight with $\NHmol > 10^{20}$ \citep[see][and references therein]{Nehme2008}.  This provides a potential basis for additional modeling work that could explore questions such as whether a concatenation model is potentially a feasible explanation for certain lines of sight.

\section{SUMMARY}
\label{s:summary}
We have undertaken a study of 22 lines of sight with properties that qualify them as ``translucent lines of sight" though not necessarily ``translucent clouds", per the distinctions discussed in the FTCS and \citet{SnowMcCall}.  We find, consistent with previous studies of the interstellar medium, that $\Hmol$ in the $J \ge 2$ states is overabundant relative to the expectation of a single, thermalized cloud.  We do not find conclusive evidence that the observed lower-$J$ and higher-$J$ $\Hmol$ are dominated by physically distinct clouds in any of these lines of sight---which we searched for in the form of trends in velocity offset and/or $b$-value as a function of $J$.  We are unable to find a convincing correlation, either positive or negative, between $b$-value and $J$-level.  Similarly, we do not see significant evidence that there are trends between velocity offsets of the $\Hmol$ transitions and $J$.   However, the results for both $b$-values and velocity offsets are subject to the systematic errors discussed in \S\ref{ss:velocitystructure}, and we are unable to state a null result more strongly.


We explored the possibility that the column densities and abundances of various atomic and molecular species are correlated with the column densities of the different $\Hmol$ $J$-levels and interesting derived ratios such as $N(4)/N(0)$.  We find many correlations with a high calculated confidence level, and virtually all of these have a straightforward physical interpretation.  These correlations exist even when accounting for the common correlation with total column density.


Further modeling is needed to understand the $J \geq 2$ $\Hmol$ in lines of sight such as these.  Our limited modeling attempts using the Meudon PDR code with two lines of sight resulted in solutions that are a poor match to several observed quantities such as the column densities of CH, CH$^+$, $^{12}$CO.  The deviation of these lines of sight from their representation as single clouds in a steady state (particularly the influence of shocks) are likely to be the main reasons for the lack of quality solutions.

\acknowledgements
We are very grateful to Dan Welty for providing access to CH and CH$^+$ velocity structure fits in addition to the other atomic and molecular column densities that are available on his website.  We also wish to thank J.~Michael Shull for access to prepublished determinations of $N(0)$ and $N(1)$ for HD 195965.  Joshua Destree and Teresa Ross provided helpful assistance with certain aspects of preparing the {\it FUSE} spectra.  Dr.~Welty and Mr.~Destree also provided many very valuable comments on the manuscript, as did the anonymous referee.  Finally, we wish to express our gratitude to the Meudon PDR code team for making their code publicly available and providing some assistance in its use.

This research was supported by an appointment to the NASA Postdoctoral Program at Goddard Space Flight Center, administered by Oak Ridge Associated Universities.  Additional support was provided by NASA grant NAG5-12279 to the University of Colorado.

\bibliographystyle{apj}
\bibliography{refs}


\clearpage \clearpage

\clearpage \clearpage

\clearpage \clearpage

\clearpage \clearpage

\clearpage \clearpage

\clearpage \clearpage

\clearpage \clearpage

\clearpage \clearpage

\clearpage \clearpage

\clearpage \clearpage

\clearpage \clearpage


\begin{deluxetable}{cccccc}
\tablecolumns{6}
\tablewidth{0pc}
\tabletypesize{\scriptsize}
\tablecaption{Lines of Sight: Stellar Data\label{stellardata}}
\tablehead{\colhead{Star Name} & \colhead{Spectral Class} & \colhead{l} & \colhead{b} & \colhead{Distance (pc)} & \colhead{Distance Ref.}}
\startdata
HD 24534 & O9.5pe & 163.08 & -17.14 & 590 & 1 \\
HD 27778 & B3V & 172.76 & -17.39 & 223 & 2 \\
HD 38087 & B5V & 207.07 & -16.26 & 480 & 1 \\
HD 40893 & B0IV & 180.09 & +4.34 & 2800 & 1 \\
HD 46056 & O8V & 206.34 & -2.25 & 2300 & 1 \\
HD 46202 & O9V & 206.31 & -2.00 & 2000 & 1 \\
HD 53367 & BOIV:e & 223.71 & -1.90 & 780 & 1 \\
HD 147888 & B3/B4V & 353.65 & +17.71 & 136 & 2 \\
HD 149404 & O9Ia & 340.54 & +3.01 & 820 & 1 \\
HD 170740 & B2V & 21.06 & -0.53 & 213 & 2 \\
HD 179406 & B3V & 28.23 & -8.31 & 160 & 1 \\
HD 185418 & B0.5V & 53.60 & -2.17 & 950 & 1 \\
HD 186994 & BOIII & 78.62 & +10.06 & 2500 & 1 \\
HD 192639 & O7.5IIIF & 74.90 & +1.48 & 1100 & 1 \\
HD 195965 & B0V & 85.71 & +5.00 & 1300 & 1 \\
HD 197512 & B1V & 87.89 & +4.63 & \nodata & \nodata \\
HD 199579 & O6.5III & 85.70 & -0.30 & 1100 & 3 \\
HD 203938 & B0.5IV & 90.56 & -2.23 & 700 & 1 \\
HD 206267 & O6(F) & 99.29 & +3.74 & 1000 & 4 \\
HD 207198 & O9II & 103.14 & +6.99 & 1000 & 1 \\
HD 207538 & O9.5V & 101.60 & +4.67 & 880 & 1 \\
HD 210839 & O6e & 103.83 & +2.61 & 505 & 2 \\
\enddata
\tablerefs{(1) From the DIBS database, http://dib.uiuc.edu/, values compiled by L.~M.~Hobbs, spectroscopic distance modulus.  (2) Hipparcos value of 4-$\sigma$ significance or better.  (3) \citet{Savage1985}.  (4) From the DIBS database, http://dib.uiuc.edu/, values compiled by L.~M.~Hobbs, stellar association.}
\end{deluxetable}

\clearpage \clearpage

\begin{deluxetable}{ccccccc}
\tablecolumns{7}
\tablewidth{0pc}
\tabletypesize{\scriptsize}
\tablecaption{Lines of Sight: Reddening Data\label{reddeningtable}}
\tablehead{\colhead{Sightline} & \colhead{$E_{B-V}$} & \colhead{Ref.} & \colhead{$A_V$} & \colhead{Ref.} & \colhead{$R_V$} & \colhead{Ref.}}
\startdata
HD 24534 & 0.59 & 1 & 2.05 & 2 & 3.47 & 2 \\
HD 27778 & 0.37 & 3 & 1.01 & 2 & 2.72 & 2 \\
HD 38087 & 0.29 & 3 & 1.61 & 4 & 5.57 & 4 \\
HD 40893 & 0.46 & 3 & 1.13 & 4 & 2.46 & 4 \\
HD 46056 & 0.50 & 3 & 1.30 & 4 & 2.60 & 4 \\
HD 46202 & 0.49 & 3 & 1.39 & 4 & 2.83 & 4 \\
HD 53367 & 0.74 & 3 & 1.76 & 4 & 2.38 & 4 \\
HD 147888 & 0.47 & 3 & 1.91 & 4 & 4.06 & 4 \\
HD 149404 & 0.62 & 5 & 2.19 & 5 & 3.53 & 5 \\
HD 170740 & 0.48 & 3 & 1.30 & 2 & 2.71 & 2 \\
HD 179406 & 0.33 & 3 & 0.94 & 4 & 2.86 & 4 \\
HD 185418 & 0.50 & 3 & 1.16 & 2 & 2.32 & 2 \\
HD 186994 & 0.17 & 3 & 0.53 & 6 & 3.10 & 7 \\
HD 192639 & 0.66 & 1 & 1.87 & 2 & 2.84 & 2 \\
HD 195965 & 0.25 & 1 & 0.77 & 8 & 3.08 & 6 \\
HD 197512 & 0.32 & 9 & 0.75 & 2 & 2.35 & 2 \\
HD 199579 & 0.37 & 9 & 1.09 & 2 & 2.95 & 2 \\
HD 203938 & 0.74 & 3 & 2.15 & 2 & 2.91 & 2 \\
HD 206267 & 0.53 & 3 & 1.41 & 2 & 2.67 & 2 \\
HD 207198 & 0.62 & 3 & 1.50 & 2 & 2.42 & 2 \\
HD 207538 & 0.64 & 3 & 1.44 & 2 & 2.25 & 2 \\
HD 210839 & 0.57 & 1 & 1.58 & 2 & 2.78 & 2 \\
\enddata
\tablerefs{(1) \citet{DS1994}.  (2) \citet{Rachford2002}; if $\av$, reference 6 also applies.  (3) From the DIBS database, http://dib.uiuc.edu/, values by the DIBS team.  (4) \citet{Rachford2009}; if $\av$, reference 6 also applies.  (5) \citet{Valencic2004}.  (6) Calculated through the definition $\rv \equiv \av / \ebv$ using values from the other two columns.  (7) $\rv$ not independently measured and is assumed to be the Galactic average of 3.1.  (8) \citet{Neckel1980}.  (9) \citet{Aiello1988}.}
\end{deluxetable}

\clearpage \clearpage

\begin{deluxetable}{cccccccccccccccccccccccc}
\tablecolumns{24}
\tablewidth{0pc}
\tabletypesize{\tiny}
\rotate
\tablecaption{Measured Equivalent Widths\label{eqwidths}}
\tablehead{\colhead{} & \colhead{} & \multicolumn{22}{c}{Line of Sight} \\ \colhead{Wavelength} & \colhead{$J$ Level} & \colhead{HD 24534} & \colhead{HD 27778} & \colhead{HD 38087} & \colhead{HD 40893} & \colhead{HD 46056} & \colhead{HD 46202} & \colhead{HD 53367} & \colhead{HD 147888} & \colhead{HD 149404} & \colhead{HD 170740} & \colhead{HD 179406} & \colhead{HD 185418} & \colhead{HD 186994} & \colhead{HD 192639} & \colhead{HD 195965} & \colhead{HD 197512} & \colhead{HD 199579} & \colhead{HD 203938} & \colhead{HD 206267} & \colhead{HD 207198} & \colhead{HD 207538} & \colhead{HD 210839}}
\startdata
919.048 & 4 & \nodata & \nodata & \nodata & \nodata & \nodata & \nodata & \nodata & \nodata & \nodata & \nodata & \nodata & \nodata & \nodata & \nodata & $83.6\pm29.7$ & \nodata & \nodata & \nodata & \nodata & \nodata & \nodata & \nodata \\
920.241 & 2 & \nodata & \nodata & \nodata & $118.8\pm53.5$ & \nodata & $128.2\pm28.1$ & \nodata & \nodata & $166.4\pm58.0$ & \nodata & \nodata & $104.4\pm29.6$ & $62.0\pm6.5$ & \nodata & $95.4\pm11.8$ & $163.1\pm50.3$ & $149.3\pm42.5$ & \nodata & \nodata & \nodata & \nodata & \nodata \\
921.730 & 3 & \nodata & \nodata & \nodata & \nodata & \nodata & \nodata & \nodata & \nodata & \nodata & \nodata & \nodata & $87.3\pm25.5$ & \nodata & \nodata & \nodata & \nodata & \nodata & \nodata & \nodata & \nodata & \nodata & \nodata \\
922.303 & 4 & \nodata & \nodata & \nodata & \nodata & \nodata & \nodata & \nodata & \nodata & \nodata & \nodata & \nodata & \nodata & \nodata & \nodata & $42.4\pm5.9$ & \nodata & \nodata & \nodata & \nodata & \nodata & \nodata & \nodata \\
922.893 & 3 & \nodata & \nodata & \nodata & \nodata & \nodata & \nodata & \nodata & \nodata & \nodata & \nodata & \nodata & \nodata & \nodata & \nodata & $104.2\pm8.9$ & \nodata & \nodata & \nodata & \nodata & \nodata & \nodata & \nodata \\
927.017 & 2 & $46.1\pm12.2$ & \nodata & \nodata & $166.0\pm19.0$ & \nodata & $161.3\pm43.9$ & \nodata & \nodata & $136.5\pm17.3$ & \nodata & \nodata & $107.5\pm26.0$ & $67.8\pm6.5$ & $107.8\pm27.3$ & $114.6\pm9.5$ & $150.7\pm38.3$ & $170.6\pm31.9$ & \nodata & $127.0\pm28.5$ & \nodata & $143.6\pm51.6$ & $142.7\pm9.3$ \\
928.436 & 3 & \nodata & \nodata & \nodata & $142.1\pm23.6$ & $114.3\pm44.3$ & $97.3\pm36.9$ & \nodata & \nodata & $128.7\pm33.7$ & \nodata & \nodata & $51.4\pm21.4$ & $73.7\pm7.8$ & $62.6\pm60.4$ & $99.7\pm6.9$ & $101.1\pm68.8$ & $149.2\pm24.1$ & \nodata & \nodata & \nodata & \nodata & $109.5\pm21.5$ \\
\enddata
\end{deluxetable}

\clearpage \clearpage

\begin{deluxetable}{ccccccccccc}
\tablecolumns{11}
\tablewidth{0pc}
\tabletypesize{\tiny}
\rotate{}
\tablecaption{Column Density Results\label{coldensities}}
\tablehead{\colhead{Sightline} & \colhead{$N$($J=0$)} & \colhead{$N$($J=1$)} & \colhead{$N$($J=2$)} & \colhead{$N$($J=3$)} & \colhead{$N$($J=4$)} & \colhead{$N$($J=5$)} & \colhead{$N$($J=6$)} & \colhead{$N$($J=7$)} & \colhead{$N$($J=8$)} & \colhead{$b$} }
\startdata
HD 24534 & $20.76\pm0.03$ & $20.42\pm0.06$ & $18.40\pm0.01$ & $17.07\pm0.03$ & $15.20^{+0.05}_{-0.04}$ & $14.21\pm0.07$ & \nodata &\nodata &\nodata & $2.6\pm0.1$ \\
HD 27778 & $20.64\pm0.05$ & $20.27\pm0.10$ & $18.47\pm0.02$ & $17.55^{+0.04}_{-0.03}$ & $15.64^{+0.06}_{-0.07}$ & $14.33^{+0.17}_{-0.19}$ & \nodata &\nodata &\nodata & $3.0^{+0.2}_{-0.1}$ \\
HD 38087 & $20.39\pm0.08$ & $20.29\pm0.05$ & $18.98\pm0.02$ & $18.24^{+0.03}_{-0.02}$ & $17.25\pm0.05$ & $16.96^{+0.10}_{-0.09}$ & $15.46^{+0.16}_{-0.15}$ & $14.89^{+0.12}_{-0.11}$ & $13.80^{+0.16}_{-0.20}$ & $2.4\pm0.1$ \\
HD 40893 & $20.27\pm0.05$ & $20.28\pm0.05$ & $18.01\pm0.05$ & $17.34^{+0.06}_{-0.07}$ & $15.21\pm0.03$ & $14.52^{+0.05}_{-0.04}$ & \nodata &\nodata &\nodata & $9.6\pm0.2$ \\
HD 46056 & $20.40\pm0.06$ & $20.35\pm0.06$ & $18.28^{+0.03}_{-0.04}$ & $17.40^{+0.06}_{-0.05}$ & $15.75^{+0.05}_{-0.06}$ & $15.06^{+0.04}_{-0.07}$ & $14.11^{+0.39}_{-1.12}$ & \nodata &\nodata & $6.2^{+0.1}_{-0.2}$ \\
HD 46202 & $20.38\pm0.07$ & $20.38\pm0.07$ & $18.40^{+0.03}_{-0.04}$ & $17.66^{+0.05}_{-0.04}$ & $15.72\pm0.03$ & $15.14^{+0.04}_{-0.05}$ & \nodata &\nodata &\nodata & $7.2\pm0.1$ \\
HD 53367 & $20.89\pm0.04$ & $20.52\pm0.07$ & $19.22\pm0.02$ & $17.66^{+0.04}_{-0.05}$ & $16.25^{+0.07}_{-0.06}$ & $15.35^{+0.08}_{-0.09}$ & $14.08^{+0.19}_{-0.24}$ & \nodata &\nodata & $4.6\pm0.1$ \\
HD 147888 & $20.39\pm0.04$ & $19.71\pm0.10$ & $18.51^{+0.02}_{-0.01}$ & $17.11\pm0.05$ & $15.65^{+0.06}_{-0.07}$ & $15.13^{+0.09}_{-0.07}$ & $14.24^{+0.16}_{-0.18}$ & \nodata &\nodata & $3.1\pm0.1$ \\
\enddata
\end{deluxetable}


\clearpage \clearpage

\begin{deluxetable}{ccccccc}
\tablecolumns{7}
\tablewidth{0pc}
\tabletypesize{\scriptsize}
\tablecaption{Limiting Column Densities\label{limits}}
\tablehead{\colhead{Sightline} & \multicolumn{4}{c}{First Undetected $J$-level} & \multicolumn{2}{c}{$J=0$, $\nu=1$} \\ \colhead{} & \colhead{$J$} & \colhead{Line (\AA{})} & \colhead{$\eqw$ (m\AA{})\tablenotemark{a}} & \colhead{$\log{N}$\tablenotemark{b}} & \colhead{$\eqw$ (m\AA{})\tablenotemark{a,c}} & \colhead{$\log{N}$\tablenotemark{b}}}
\startdata
HD 24534 & 6 & 1021.21 & 2.85 & 13.68 & 1.15 & 12.48 \\ 
HD 27778 & 6 & 1054.00 & 2.29 & 13.73 & 3.18 & 12.98 \\ 
HD 38087 & 9 & 1034.29 & 3.37 & 13.78 & \multicolumn{2}{c}{Detected\tablenotemark{d}} \\ 
HD 40893 & 6 & 1021.21 & 3.32 & 13.68 & 2.35 & 12.78 \\ 
HD 46056 & 7 & 1028.80 & 7.07 & 14.09 & 3.60 & 12.99 \\ 
HD 46202 & 6 & 1054.00 & 3.30 & 13.87 & 3.96 & 13.03 \\ 
HD 53367 & 7 & 1028.80 & 7.26 & 14.15 & 4.00 & 13.06 \\ 
HD 147888 & 7 & 1028.80 & 4.66 & 13.95 & 1.63 & 12.64 \\
\enddata
\tablenotetext{a}{This is the 1-$\sigma$ limit on $\eqw$}
\tablenotetext{b}{A 3-$\sigma$ limit on the column density.  1-$\sigma$ limits on $\eqw$ are multiplied by 3 (per Eq.~\ref{eq:eqwlimit}), and then that upper limit is matched to the curve of growth $b$-value from Table \ref{coldensities}.}
\tablenotetext{c}{Based on a line at 1052.73 \AA{}}
\tablenotetext{d}{See Table \ref{vibrational}}
\end{deluxetable}

\clearpage \clearpage

\begin{deluxetable}{cccccc}
\tablecolumns{6}
\tablewidth{0pc}
\tabletypesize{\small}
\tablecaption{Vibrationally Excited Column Densities\label{vibrational}}
\tablehead{\colhead{Sightline} & \colhead{$J$} & \colhead{$\nu$} & \colhead{Line (m\AA{})} & \colhead{$\eqw$} & \colhead{$\log{N}$\tablenotemark{a}}}
\startdata
HD 38087 & 0 & 1 & 1052.73 & $16.26\pm5.02$ & $13.36^{+0.26}_{-0.25}$ \\
		 & 1 & 1 & 1052.61 & $13.55\pm5.58$ & $13.56^{+0.28}_{-0.32}$ \\
		 & 1 & 1 & 1097.86 & $4.37\pm2.63$   & $13.27^{+0.24}_{-0.43}$ \\
		 & 1 & 1 & 1128.22 & $18.06\pm5.44$ & $14.00^{+0.27}_{-0.25}$ \\
\multicolumn{3}{c}{Alternate fit\tablenotemark{b}} & & & $13.54^{+0.11}_{-0.15}$ \\
\multicolumn{3}{c}{Alternate fit\tablenotemark{c}} & & & $13.45^{+0.31}_{-0.21}$ \\
HD 199579 & 0 & 1 & 1052.73 & $5.04\pm1.84$ & $12.62^{+0.14}_{-0.20}$ \\
\enddata
\tablenotetext{a}{$J=0$, $\nu=1$ column densities are calculated by fitting the equivalent width to the corresponding curve-of-growth with the $b$-value from the solution in Table \ref{coldensities}, save for the exceptions noted.}
\tablenotetext{b}{An alternate solution of the $J=1$, $\nu=1$ column density for HD 38087 based on a curve of growth fit assuming the $b$-value from Table \ref{coldensities} for all lines simultaneously instead of individually.}
\tablenotetext{c}{An alternate solution of the $J=1$, $\nu=1$ column density for HD 38087 based on an independent curve of growth fit for the three measured lines.  The fit for the $b$-value is $30\kmpers$, which should be interpreted to mean that the best fit places all lines on the linear portion of the curve of growth (see discussion in \S\ref{sss:indbvalues}).}
\end{deluxetable}

\clearpage \clearpage
\begin{deluxetable}{ccccccccc}
\tablecolumns{9}
\tablewidth{0pc}
\tabletypesize{\tiny}
\tablecaption{Correlations with Other Species---Column Densities\label{corrtablecols}\tablenotemark{a}}
\tablehead{\colhead{Element} & \colhead{$\NHmol$, total} & \colhead{$N(2)/N(0)$} & \colhead{$N(3)/N(1)$} & \colhead{$N(4)/N(0)$} & \colhead{$N(5)/N(1)$} & \colhead{$N(4)/N(2)$} & \colhead{$N(5)/N(3)$} & \colhead{$N(6)/N(2)$}}
\startdata
Li I & -0.477 & -0.065 & -0.046 & -0.002 & 0.217 & 0.049 & 0.321 & -0.125 \\
 & 83.7\% & 14.2\% & 10.1\% & 0.5\% & 45.3\% & 10.7\% & 63.4\% & 21.1\% \\
 & 10 & 10 & 10 & 10 & 10 & 10 & 10 & 7 \\
K I & 0.504 & -0.542 & -0.651 & -0.559 & -0.525 & -0.308 & -0.174 & -0.502 \\
 & 95.4\% & 97.0\% & 99.4\% & 97.6\% & 96.3\% & 75.4\% & 48.2\% & 86.1\% \\
 & 16 & 16 & 16 & 16 & 16 & 16 & 16 & 10 \\
Ca I & -0.133 & -0.227 & -0.192 & -0.135 & -0.041 & 0.016 & 0.100 & 0.187 \\
 & 37.7\% & 60.3\% & 52.5\% & 38.2\% & 12.0\% & 4.8\% & 28.7\% & 41.7\% \\
\enddata
\tablenotetext{a}{For each element there are three rows.  The first row is the Pearson correlation coefficient $r$.  The second row is the confidence level that $r \ne 0$ for a $t$-distribution of the quantity $r(n-2)^{1/2}(1-r^2)^{-1/2}$ for $n-2$ degrees of freedom, where $n$ is the number of data points.  The third row is $n$.  Combinations with fewer than three available data points are omitted.}
\end{deluxetable}

\clearpage \clearpage

\begin{deluxetable}{ccccccccc}
\tablecolumns{9}
\tablewidth{0pc}
\tabletypesize{\tiny}
\tablecaption{Correlations with Other Species---Other Elemental Ratios\label{corrtableratios}\tablenotemark{a}}
\tablehead{\colhead{Element} & \colhead{$\NHmol$, total} & \colhead{$N(2)/N(0)$} & \colhead{$N(3)/N(1)$} & \colhead{$N(4)/N(0)$} & \colhead{$N(5)/N(1)$} & \colhead{$N(4)/N(2)$} & \colhead{$N(5)/N(3)$} & \colhead{$N(6)/N(2)$}}
\startdata
CH$^+$/CH & -0.311 & 0.433 & 0.397 & 0.660 & 0.621 & 0.577 & 0.445 & 0.461 \\
 & 80.5\% & 93.6\% & 90.8\% & 99.8\% & 99.5\% & 99.0\% & 94.4\% & 86.9\% \\
 & 19 & 19 & 19 & 19 & 19 & 19 & 19 & 12 \\
CH$^+$/CN & -0.321 & 0.346 & 0.132 & 0.381 & 0.365 & 0.193 & 0.284 & 0.002 \\
 & 79.1\% & 82.6\% & 38.7\% & 86.8\% & 85.0\% & 54.2\% & 73.1\% & 0.5\% \\
 & 17 & 17 & 17 & 17 & 17 & 17 & 17 & 11 \\
Fe II/Fe I & 0.148 & 0.167 & -0.074 & 0.401 & 0.109 & 0.397 & 0.203 & 0.353 \\
 & 33.6\% & 37.5\% & 17.1\% & 77.8\% & 25.1\% & 77.3\% & 45.0\% & 61.0\% \\
 & 11 & 11 & 11 & 11 & 11 & 11 & 11 & 8 \\

\enddata
\tablenotetext{a}{See notes on Table \ref{corrtablecols}.}
\end{deluxetable}

\clearpage \clearpage

\begin{deluxetable}{ccccccccc}
\tablecolumns{9}
\tablewidth{0pc}
\tabletypesize{\tiny}
\tablecaption{Correlations with Other Species---$J$ Levels\label{corrtableJ}\tablenotemark{a}}
\tablehead{\colhead{Element} & \colhead{$J=0$} & \colhead{$J=1$} & \colhead{$J=2$} & \colhead{$J=3$} & \colhead{$J=4$} & \colhead{$J=5$} & \colhead{$J=6$} & \colhead{$J=7$}}
\startdata
Li I & -0.531 & -0.033 & -0.612 & -0.369 & -0.206 & -0.120 & -0.768 & \nodata \\
 & 85.9\% & 6.7\% & 92.0\% & 67.2\% & 40.5\% & 24.1\% & 92.5\% & \nodata \\
 & 10 & 10 & 10 & 10 & 10 & 10 & 7 & \nodata \\
K I & 0.430 & 0.474 & -0.444 & -0.550 & -0.384 & -0.441 & -0.732 & -0.929 \\
 & 89.1\% & 92.6\% & 90.3\% & 96.7\% & 84.2\% & 90.0\% & 97.5\% & 75.9\% \\
 & 16 & 16 & 16 & 16 & 16 & 16 & 10 & 4 \\
Ca I & -0.377 & 0.157 & -0.588 & -0.268 & -0.292 & -0.068 & -0.347 & \nodata \\
 & 83.4\% & 42.3\% & 97.9\% & 66.7\% & 70.9\% & 18.9\% & 67.3\% & \nodata \\
\enddata
\tablenotetext{a}{See notes on Table \ref{corrtablecols}.  The correlation coefficients in this table are partial correlation coefficients that assume $\NHtot$ is held fixed to reduce the effect of spurious correlations arising from total column density effects.}
\end{deluxetable}

\clearpage \clearpage

\begin{deluxetable}{cccc}
\tablecolumns{4}
\tablewidth{0pc}
\tablecaption{Excitation Temperatures\label{texctable}}
\tabletypesize{\small}
\tablehead{\colhead{Sightline} & \colhead{$T_{01}$\tablenotemark{a}} & \colhead{$T_{02}$\tablenotemark{b}} & \colhead{$T_{3+}$\tablenotemark{c}}}
\startdata
HD 24534 & $57\pm3$ & $73\pm1$ & $211\pm5$ \\
HD 27778 & $55\pm5$ & $78\pm1$ & $192\pm7$ \\
HD 38087 & $70\pm6$ & $120\pm4$ & $460\pm11$ \\
HD 40893 & $78\pm6$ & $74\pm2$ & $237\pm6$ \\
HD 46056 & $73\pm6$ & $79\pm2$ & $263\pm8$ \\
HD 46202 & $77\pm8$ & $83\pm2$ & $247\pm6$ \\
HD 53367 & $55\pm3$ & $96\pm2$ & $282\pm10$ \\
HD 147888 & $45\pm3$ & $88\pm1$ & $340\pm13$ \\
HD 149404 & $60\pm3$ & $72\pm1$ & $406\pm9$ \\
HD 170740 & $71\pm8$ & $91\pm2$ & $509\pm34$ \\
HD 179406 & $59\pm5$ & $67\pm2$ & $342\pm17$ \\
HD 185418 & $100\pm9$ & $79\pm1$ & $220\pm4$ \\
HD 186994 & $96\pm8$ & $93\pm2$ & $215\pm3$ \\
HD 192639 & $97\pm9$ & $85\pm2$ & $300\pm7$ \\
HD 195965 & $109\pm7$ & $90\pm2$ & $262\pm3$ \\
HD 197512 & $94\pm8$ & $78\pm2$ & $328\pm13$ \\
HD 199579 & $69\pm3$ & $76\pm1$ & $439\pm4$ \\
HD 203938 & $74\pm7$ & $91\pm2$ & $302\pm13$ \\
HD 206267 & $64\pm3$ & $68\pm1$ & $336\pm7$ \\
HD 207198 & $65\pm3$ & $73\pm1$ & $284\pm8$ \\
HD 207538 & $72\pm6$ & $81\pm2$ & $225\pm5$ \\
HD 210839 & $72\pm4$ & $70\pm1$ & $545\pm8$ \\
\enddata
\tablenotetext{a}{Temperature derived through linear fit of $\log{N_J/g_J}$ for $J =0-1$, column densities from \citet{Rachford2002,Rachford2009}.}
\tablenotetext{b}{Temperature derived through linear fit of $\log{N_J/g_J}$ for $J=0-2$.}
\tablenotetext{c}{Temperature derived through linear fit of $\log{N_J/g_J}$ for $J \geq 3$.}
\end{deluxetable}

\clearpage \clearpage

\begin{deluxetable}{ccc}
\tablecolumns{3}
\tablewidth{0pc}
\tablecaption{Model Input Parameters\label{modelinput}\tablenotemark{a}}
\tablehead{\colhead{Parameter} & \colhead{HD 46056} & \colhead{HD 147888\tablenotemark{b}}}
\startdata
$\av$ & 1.30 & 1.94 \\
$\ebv$ & 0.49 & 0.47 \\
$\rv$ & 2.66 & 4.13 \\
y$_0$ & 4.581 & 4.587 \\
$\gamma$ & 0.911 & 1.022 \\
c$_1$ & -0.641 & 1.611 \\
c$_2$ & 0.878 & 0.133 \\
c$_3$ & 3.559 & 3.823 \\
c$_4$ & 0.609 & 0.339 \\
$\NHtot$ & $3.4\times10^{21}$ & $5.7\times10^{21}$ \\
$\NHtot / \ebv$ & $6.9\times10^{21}$ & $1.2\times10^{22}$ \\
\enddata
\tablenotetext{a}{Extinction curve parameters are from \citet{Valencic2004}.  $\NHtot$ is from Lyman-$\alpha$ measurements:  \citet{DS1994} for HD 46056 and \citet{Cartledge2004} for HD 147888 (note the extensive discussion in that paper due to concerns about stellar contamination).}
\tablenotetext{b}{Slight discrepancies exist for $\av$ and $\rv$ for HD 147888 between this table and Table \ref{reddeningtable} due to differences in \citet{Rachford2002, Rachford2009} and \citet{Valencic2004}---\citet{Rachford2002, Rachford2009} values were preferred for Table \ref{reddeningtable}, but the \citet{Valencic2004} values were the published values available at the time the models were calculated.}
\end{deluxetable}

\clearpage \clearpage

\begin{deluxetable}{lcccc}
\tablecolumns{5}
\tablewidth{0pc}
\tabletypesize{\tiny}
\tablecaption{Modeling Output Parameters\label{modeloutput}}
\tablehead{\colhead{} & \multicolumn{2}{c}{HD 46056} & \multicolumn{2}{c}{HD 147888} \\ \colhead{Parameter} & \colhead{Observations\tablenotemark{a}} & \colhead{Best Model Fit} & \colhead{Observations\tablenotemark{a}} & \colhead{Best Model Fit}}
\startdata
$N(0)$ & $20.40\pm0.06$                & 20.99 & $20.39\pm0.04$               & 21.28 \\
$N(1)$ & $20.35\pm0.06$                & 20.80 & $19.71\pm0.10$               & 20.91 \\
$N(2)$ & $18.28^{+0.03}_{-0.04}$ & 18.88 & $18.51^{+0.02}_{-0.01}$ & 18.45 \\
$N(3)$ & $17.40^{+0.06}_{-0.05}$ & 17.65 & $17.11\pm0.05$                &16.68 \\
$N(4)$ & $15.75^{+0.05}_{-0.06}$ & 15.89 & $15.65^{+0.06}_{-0.07}$ &15.72 \\
$N(5)$ & $15.06^{+0.04}_{-0.07}$ & 15.48 & $15.13^{+0.09}_{-0.07}$ & 15.26 \\
$N(6)$ & $14.11^{+0.39}_{-1.12}$ & 14.52 & $14.24^{+0.16}_{-0.18}$ & 14.33 \\
$\log{N(\rm{CH})}$                                     & $13.31\pm0.06$   & 14.28 & $13.30\pm0.02$ & 14.59 \\
\enddata
\tablenotetext{a}{Observed column densities and subsequently derived quantities.  $\Hmol$ from this paper, other column densities from http://www.astro.uiuc.edu/~dwelty/coldens.html.}
\end{deluxetable}

\clearpage \clearpage

\end{document}

%% file: macros.tex
\newcommand{\kmpers}{{{\rm \; km\;s}^{-1}}}

\newcommand{\percc}{{\rm \; cm}^{-3}}
\newcommand{\dex}{{\rm \; dex}}
\newcommand{\magnitude}{{\rm \; mag}}

\newcommand{\NHI}{{N{\rm(H \; I)}}}

\newcommand{\NHmol}{{N\rm(H_{2})}}

\newcommand{\NHtot}{{N{\rm(H_{tot})}}}
\newcommand{\Hmol}{{\rm H_{2}}}

\newcommand{\nHvol}{{\rm n_{H}}}

\newcommand{\fHmol}{{f\rm(H_{2})}}
\newcommand{\eqw}{{W_{\lambda}}}
\newcommand{\av}{{A_V}}

\newcommand{\ebv}{{E_{B-V}}}

\newcommand{\rv}{{R_V}}